\documentclass[a4paper]{jpconf}
\usepackage{amsmath, amssymb}
\usepackage{amsfonts}
\usepackage{xcolor}
\usepackage{graphicx}
\usepackage{rotating}
\usepackage{colortbl}
\usepackage{subfig}
\usepackage{caption}

\renewcommand{\imath}[0]{\mathrm{i}}
	
\newcommand{\opn}[1]{\mathrm{#1}}
\renewcommand{\bs}[1]{\boldsymbol{#1}}
\newcommand{\ten}[1]{\boldsymbol{\mathcal{#1}}}

\begin{document}

\title{Radiative heat transfer between spatially nonlocally responding 
dielectric objects}
\author{Robin Schmidt and Stefan Scheel}

\address{Institut f\"ur Physik, Universit\"at Rostock, Albert-Einstein-Stra{\ss}e 23, D-18059 Rostock, Germany}

\ead{robin.schmidt@uni-rostock.de}

\begin{abstract}
We calculate numerically the heat transfer rate between a spatially dispersive
sphere and a half-space. By utilising Huygens' principle and the extinction 
theorem, we derive the necessary reflection coefficients at the sphere and the 
plate without the need to resort to additional boundary conditions. We find for 
small distances $d\sim 1$nm a significant modification of the spectral heat 
transfer rate due to spatial dispersion. As a consequence, the spurious 
divergencies that occur in spatially local approach are absent. 
\end{abstract}

\section{Introduction}
In the past decade, the advances in nanooptics and nanophotonics made the 
fabrication of ever smaller dielectric and metallic objects feasible, whose 
separation can be controlled down to a few nanometers
\cite{Kim2015,Shen2009,Rousseau2009,Song2015}.  The near-field of such 
nanoscopic objects exhibit strong mode confinement and field enhancement due to 
the presence of evanescent fields. This results in an increase of heat transfer 
rates several orders of magnitude larger than the classical far-field limit.  
These enhancements open up new possibilities in numerous applications including 
thermal near-field imaging \cite{Kittel2005}, heat-assisted magnetic recording 
\cite{Challener2009}, nanopatterning \cite{Lee2008}, and near-field 
thermophotovoltaics \cite{Basu2009}.

It is well known that thermal heat transfer described by black-body radiation 
is an insufficient description for separation distances smaller than the 
thermal wavelength $\lambda_T$. It strongly underestimates the near-field 
contribution from photon tunnelling mediated by surface phonon-polaritons in 
polar dielectrics or (spoof) surface plasmon-polaritons at metal interfaces 
\cite{Dai2015,Pendry2004} or doped silicon \cite{Fu2006} in the near-IR.  
A theoretical framework, namely fluctuation electrodynamics, has been developed 
by Rytov as early as the 1950's (see, e.g. the textbook \cite{Rytov1987}). 
In this theory,  electromagnetic radiation is created by either the thermal 
random motion of charge carries, i.e.  fluctuating current densities, or 
fluctuation dipoles in polar media. In linear response theory, these 
fluctuations are ultimately linked, via the fluctuation-dissipation theorem, to 
their respective response functions (e.g. conductivity or dielectric 
permittivity). In the commonly employed spatially local limit, the near-field 
heat transfer varies with gap distance $d$ for the respective geometries, as 
$d^{-2}$ (plate-plate), $ d^{-1.5}$ (cylinder–cylinder) and $d^{-1}$ 
(sphere-plate). These findings agree well with several recent experiments 
\cite{Rousseau2009, Song2015, St-Gelais2016, Song2016, Chen2015}.

However, in the spatially local theory the calculated heat transfer rate 
diverges as soon as the separation of the bodies vanishes, i.e. when 
$d\rightarrow 0$.
Recent works \cite{Philbin2014,Singer2014, Henkel2006,Chapuis2008} 
showed that, in order to overcome these divergences, spatial nonlocality (or 
spatial dispersion) of the dielectric tensor must be taken into account.  In 
addition, spatial dispersion changes the mode structure of surface plasmons 
and thus the spectral heat transfer rate. 
In general, spatial nonlocality becomes important whenever the electromagnetc 
field varies appreciably on a length scale comparable with the intrinsic 
natural length scales of a medium, e.g. the electron mean free path or 
interatomic spacings.  Well-known effects associated with spatial dispersion 
are the anomalous skin effect and resonant excitation of longitudinal modes in 
nanoparticles \cite{Melrose1991}, band-gap photoluminescence and dynamical 
screening \cite{Ropke1998}.

When considering the transition of electromagnetic waves between two media one 
usually fits the bulk modes at the interface according Maxwell boundary 
conditions (MBCs).  However, in a spatially dispersive media there exist 
additional longitudinal and transverse modes by virtue of the dispersion 
relations
\begin{gather}
k^2=k^2_0\varepsilon_{\bot}(\mathbf{k},\omega), \quad 
\varepsilon_{\parallel}(\mathbf{k},\omega)=0\:,\label{dispRel1}
\end{gather}
where $k_0=\omega/c$ is the wave number of free space, 
$\varepsilon_{\bot}(\mathbf{k},\omega)$ is the transverse dielectric function 
and $\varepsilon_{\parallel}(\mathbf{k},\omega)$ is the longitudinal dielectric 
function. At the interface of a medium with free space, these modes have to be 
match with the free-space modes for which the Maxwell boundary conditions are 
seemingly insufficient. Pekar \cite{Pekar1958} and later Hopfield and 
Thomas \cite{Hopfield1963} provided the missing relations by introducing the
\textit{ad hoc} concept of additional boundary conditions (ABCs). Since then, 
the applicability of these and other ABCs \cite{bookHalevi1992} have been a 
matter of controversy \cite{Henneberger1998,Agarwal1975,Muljarov2002}. A 
disadvantage of ABCs is that the number ABC's required to match the modes at the 
surface depends on the number of supported modes by the media. In addition, the 
choice of the appropriate ABC depends on the particular system under 
consideration \cite{Maslovski2010}. However, when applying the extinction 
theorem together with Huygens' principle, Maxwell's boundary conditions are 
indeed sufficient as they contain all information required to match any number 
of modes for any spatially dispersive media \cite{Schmidt2016}.

In this article, we investigate the heat transfer rate between a sphere and a 
half-space of different types of spatially dispersive media separated by a 
vacuum gap. We follow the macroscopic approach of fluctuation electrodynamics in 
which the sources of thermal radiation are expressed via the 
fluctuation-dissipation theorem.  The boundary value problem is dealt with by 
virtue of Huygens' principle and the extinction theorem (HuyEx).
In Sec.~\ref{sec:Ex&Huy} we apply this principle to the scattering problem at
a single plate and compare the obtained reflection coefficients with different
ABC approaches. In Sec.~\ref{sec:sphere-emission} we derive the thermal emissivity
of a spatially dispersive sphere. We combine these results in 
Sec.~\ref{sec:SpherePlate} to calculate the net heat transfer in the 
plate-sphere geometry and we present our numerical results. 
%
\section{Extinction theorem and Huygens' principle}\label{sec:Ex&Huy}
\subsection{Source-quantity representation of the electromagnetic field}

The evolution of an electromagnetic wave is described by the Helmholtz 
equation. In the case of a spatially nonlocal medium the relation between the 
displacement field $\mathbf{D}(\mathbf{r},\omega)$ and the electric field 
$\mathbf{E}(\mathbf{r},\omega)$ is given by a convolution integral
\begin{gather}
\mathbf{D}(\mathbf{r},\omega)=\int\limits\opn{d}^3r'\:
\ten{\varepsilon}(\mathbf{r},\mathbf{r}',\omega)\cdot 
\mathbf{E}(\mathbf{r}',\omega)
\end{gather}
where $\ten{\varepsilon}(\mathbf{r},\mathbf{r}',\omega)$ is the dielectric 
permittivity tensor. It depends both on the source position $\mathbf{r}'$ at 
which the field excitation occurs as well as on the observation point 
$\mathbf{r}$ which in general is different from $\mathbf{r}'$. Note that a
spatially nonlocal permittivity also partially covers (para-)magnetic media
\cite{Melrose1991}. The propagation of the electromagnetic field is 
then governed by the Helmholtz equation. For spatially dispersive media it 
becomes an integrodifferential equation, viz.
\begin{gather}
\nabla\times\nabla\times\mathbf{E}(\mathbf{r},\omega)-\frac{\omega^2}{c^2}
\int\limits\opn{d}^3r'\:\ten{\varepsilon}(\mathbf{r},\mathbf{r}',\omega)\cdot
\mathbf{E}(\mathbf{r}',\omega) =i\mu_0\omega\mathbf{j}(\mathbf{r},\omega) \:.
\label{Helmholtz}
\end{gather}

The electromagnetic field is driven by the current density 
$\mathbf{j}(\mathbf{r},\omega)$ which becomes the central quantity in 
fluctuation electrodynamics as well in electromagnetic field quantisation 
within the framework of macroscopic quantum electrodynamics 
\cite{Gruner1996,Dung1998,Scheel1998,MatloobLoudon1996,Matloob1999,Acta2008,Raabe2007}. 
The formal solution of the Helmholtz is then
\begin{gather}
\mathbf{E}(\mathbf{r},\omega) = i\mu_0\omega\int\limits_{\mathbb{R}^3}
\opn{d}^3r'\:{\ten G}(\mathbf{r},\mathbf{r}',\omega)
\cdot\mathbf{j}(\mathbf{r}',\omega)
\label{FormalE-Solution}\; .
\end{gather}
The dyadic Green function ${\ten G}(\mathbf{r},\mathbf{r}',\omega)$ is the 
fundamental solution to the Helmholtz equation (\ref{Helmholtz}) and is thus 
the unique solution to
\begin{gather}
\nabla\times\nabla\times{\ten G}(\mathbf{r},\mathbf{r}',\omega)
-\frac{\omega^2}{c^2}\int\limits\opn{d}^3s\:
\ten{\varepsilon}(\mathbf{r},\mathbf{s},\omega)
\cdot{\ten G}(\mathbf{s},\mathbf{r}',\omega) 
={\ten I}\delta(\mathbf{r}-\mathbf{r}')
\label{Helmholtz-Green}
\end{gather}
together with the appropriate boundary conditions at infinity. It describes 
the propagation of an elementary dipolar excitation from $\mathbf{r}'$ to 
$\mathbf{r}$. The dyadic Green function contains all information about 
the electromagnetic response as well as the geometries of the dielectric media 
involved. Due to the linearity of the Helmholtz equation, the dyadic Green 
function can be decomposed into a bulk part 
$\ten{G}_{\opn{i}}(\mathbf{r},\mathbf{r}',\omega)$ inside one medium, and a 
scattering part $\ten{G}^{(fs)}(\mathbf{r},\mathbf{r}',\omega)$ describing 
transmission and reflection at interfaces between media as
\begin{gather}
\ten{G}^{(fs)}(\mathbf{r},\mathbf{r}',\omega)
=\ten{G}_{\opn{i}}(\mathbf{r},\mathbf{r}',\omega)\delta_{fs}
+\ten{G}^{(fs)}(\mathbf{r},\mathbf{r}',\omega)\;.
\label{LinarGreen}
\end{gather}
Here, $s$ and $f$ denote the regions of the $s$ource and $f$ield points,
respectively. Furthermore, the dyadic Green function is reciprocal 
$\ten{G}(\mathbf{r},\mathbf{r}',\omega)=\ten{G}^T(\mathbf{r}',\mathbf{r},
\omega)$, and it is analytic in the upper half of the complex $\omega$ plane 
and obeys the Schwarz reflection principle 
$\ten{G}^\ast(\mathbf{r},\mathbf{r}',\omega)
=\ten{G}(\mathbf{r},\mathbf{r}',-\omega^\ast)$.

\subsection{The dyadic Green function for spatially dispersive bulk media}

In the case of an infinitely extended homogeneous medium, the dielectric 
permittivity becomes translationally invariant,
$\ten{\varepsilon}(\mathbf{r},\mathbf{r}',\omega)
\equiv\ten{\varepsilon}(\mathbf{r}-\mathbf{r}',\omega)$, but still includes
potential anisotropy, absorption, as well as spatial and temporal 
dispersion. The translational invariance allows one to solve 
Eq.~(\ref{Helmholtz-Green}) for the bulk Green tensor 
$\ten{G}(\mathbf{r},\mathbf{r}',\omega)$ by Fourier 
transform techniques. As mentioned above, in a spatially dispersive medium it 
is impossible to distinguish between (transverse) electric and magnetic 
responses. This allows to gauge all electromagnetic responses into a single 
dielectric permittivity tensor \cite{Melrose1991}. If the dielectric tensor 
fulfils either the relation 
$\ten{\varepsilon}(\mathbf{k},\omega)=\ten{\varepsilon}^{T}(\mathbf{k},\omega)$ 
or $\ten{\varepsilon}(\mathbf{k},\omega)=\ten{\varepsilon}(-\mathbf{k},\omega)$ 
in Fourier space, the material is nongyrotropic. 

The dielectric tensor of a homogeneous and isotropic medium with
$\ten{\varepsilon}(\mathbf{r}-\mathbf{r}',\omega)$
$=\varepsilon(\mathbf{r}-\mathbf{r}',\omega)\ten{I}$ can be decomposed into a 
transverse and a longitudinal part with respect to the wave vector 
$\mathbf{k}$. In spatial Fourier space this decomposition reads 
\begin{gather}
\epsilon(\mathbf{k},\omega)=\epsilon_{\perp}(\mathbf{k},\omega)
\bigg(\ten{I}-\frac{\mathbf{k}\otimes \mathbf{k}}{k^2}\bigg) 
+\epsilon_{\parallel}(\mathbf{k},\omega)\frac{\mathbf{k}\otimes 
\mathbf{k}}{k^2}\:,
\end{gather}
where $\epsilon_{\perp}(\mathbf{k},\omega)$ and  
$\epsilon_{\parallel}(\mathbf{k},\omega)$ are the scalar transverse and 
longitudinal dielectric functions, respectively. Solving 
Eq.~(\ref{Helmholtz-Green}), the bulk Green tensor is then given by
\begin{gather}
{\ten G}_{\opn{i}}(\mathbf{r},\mathbf{r}',\omega)= 
\int\frac{\opn{d}^3k}{(2\pi)^3} e^{i\mathbf{k}\cdot(\mathbf{r}-\mathbf{r}')} 
\bigg[\frac{\ten{I}-\mathbf{k}\otimes\mathbf{k}/k^2}{D^{(\opn{i})}_{\bot}(k,\omega)}
-\frac{\mathbf{k}\otimes\mathbf{k}/k^2}{D^{(\opn{i})}_{\|}(k,\omega)}\bigg]\;,
\label{Bulk-Green}
\end{gather}
where $D^{(\opn{i})}_{\|}(k,\omega)=k_0^2\epsilon^{(\opn{i})}_{\parallel}(k,\omega)$ 
and $D^{(\opn{i})}_{\bot}(k,\omega)=k^2-k_0^2\epsilon^{(\opn{i})}_{\bot}(k,
\omega)$ are the dispersion relations in medium $\opn{i}$ for longitudinal and 
transverse waves, respectively. 

Equation~(\ref{Bulk-Green}) is a general expression for the bulk medium Green 
function as we did not yet specify the dielectric permittivity function of the 
medium. Depending on the material under consideration, some examples of known 
spatially dielectric dispersive dielectric functions are the random-phase 
approximation (RPA) \cite{Bohm1951} used for plasmas (or metals), with the 
later extension to the Mermin \cite{Mermin1970} and Born--Mermin 
\cite{Reinholz2005} approximations. For the purpose of our numerical 
evaluation we will use the damped harmonic oscillator model \cite{Hopfield1963}
\begin{gather}
\epsilon(k_, \omega) = \epsilon_b + 
\frac{\omega_p^2}{\omega_T^2 + D k^2 - \omega^2 -i \gamma \omega}\;,
\label{Epsilon_ZnSe}
\end{gather}
with $\epsilon_b$ the background dielectric function and $\gamma$ the damping 
constant. Here, $\hbar \omega_T=E_g+E_n$ is the energy required to create a 
motionless exciton with principal quantum number $n$ and effective mass $M$. 
The dispersion parameter is given by $D=\hbar\omega_T/M$ 
\cite{Cocoletzia2005}. For the material ZnSe we will use the parameters 
$\omega_T=2.8$eV, $\gamma=10^{-5}\omega_T$, $\omega^2_p=5.5 \cdot 
10^{-3}\omega^2_T$ and $D=5.5\cdot10^{-5}c^2/\epsilon_b$ throughout this 
manuscript.

\subsection{Extinction theorem and Huygens' principle}\label{subsec:Ex&Huy}

The starting point are the expressions for the extinction theorem and Huygens' 
principle for two media separated by a single interface. Their construction is 
based on the bulk Helmholtz equation for spatially dispersive media, 
Eqs.~(\ref{Helmholtz}) and (\ref{Helmholtz-Green}), as has been shown in 
Refs.~\cite{Schmidt2016,Raabe2007, Chew1995}. They are a direct consequence of 
Maxwell's equations with the only assumptions of a sufficiently sharp boundary 
(dielectric approximation), the validity of the reciprocity relation for the 
Green tensor and the radiation condition, i.e. that all electromagnetic fields 
vanish at infinity. This set of equations reads as
\begin{gather}
\mathbf{E}_0(\mathbf{r},\omega)\Theta_{V_0}(\mathbf{r})
=\mathbf{E}_{inc}(\mathbf{r},\omega) \nonumber +i\omega\mu_0\int\limits_{\partial V}\opn{d}a(\mathbf{r}')\:
\bigg\lbrace\mu_0^{-1}{\ten G}_{0}(\mathbf{r},\mathbf{r}',\omega)\cdot
[\mathbf{n}(\mathbf{r}')\times\mathbf{B}_0(\mathbf{r}',\omega)] \nonumber \\
-\bs{\mathit\Gamma}^T_{0}(\mathbf{r}',\mathbf{r},\omega)\cdot
[\mathbf{n}(\mathbf{r}')\times\mathbf{E}_0(\mathbf{r}',\omega)]\bigg\rbrace
\label{E-Field Vacuum}
\end{gather}
and
\begin{gather}
\mathbf{E}_1(\mathbf{r},\omega)\Theta_{V_1}(\mathbf{r})= 
-i\mu_0\omega\int\limits_{\partial V}\opn{d}a(\mathbf{r}')
\:\bigg\lbrace\mu_0^{-1}{\ten G}_{1}(\mathbf{r},\mathbf{r}',\omega)\cdot
[\mathbf{n}(\mathbf{r}')\times\mathbf{B}_1(\mathbf{r}',\omega)] \nonumber \\ 
-\bs{\mathit\Gamma}^T_{1}(\mathbf{r}',\mathbf{r},\omega)\cdot
[\mathbf{n}(\mathbf{r}')\times\mathbf{E}_1(\mathbf{r}',\omega)]\bigg\rbrace
\label{E-Field Medium}\;.
\end{gather}
Here, $\mathbf{E}_{inc}$ is the incoming field and indices $0,1$ label the 
respective media. The characteristic function $\Theta_{V_i}(\mathbf{r})$ of the 
body labelled with index $i$ is one if the point $\mathbf{r}$ is inside of 
volume $ V_i$ and vanishes otherwise. The magnetic Green tensor is denoted by
\begin{equation}
\bs{\mathit{\Gamma}}_{\opn{i}}(\mathbf{r},\mathbf{r}',\omega) = -(i\mu_0\omega)^{-1} 
\bs{\nabla}\times{\ten G}_{\opn{i}}(\mathbf{r},\mathbf{r}',\omega)\:.
\end{equation}
The Green function $\ten{G}_i(\mathbf{r},\mathbf{r}',\omega)$ is the bulk Green 
function of medium $i$ and the vector $\mathbf{ n}(\mathbf{r}')$ denotes the 
unit normal vector at the interface location $\mathbf{r}'$. 
Both Eqs.~(\ref{E-Field Vacuum}) and (\ref{E-Field  Medium}) are connected via 
the Maxwell boundary conditions (MBC)
\begin{gather}
\mathbf{ n}\times \mathbf{ B}_0=\mathbf{ n}\times \mathbf{ B}_1\:,\qquad \mathbf{ n}\times \mathbf{ 
E}_0=\mathbf{ n}\times \mathbf{E}_1\label{MBC}\;
\end{gather}
at the interface.

At this point, let us remark on the chosen boundary conditions, and how these
fit the additional amplitudes for which ABC's are usually required. By virtue 
of the source quantity representation, Eq.~(\ref{FormalE-Solution}), it becomes
evident that once the source current density is known, one is able to compute 
the electric field components for every possible number of bulk modes uniquely. 
Suppose that a plane electromagnetic wave propagates, e.g. along the 
$-\mathbf{e}_z$ direction in an infinitely extended bulk medium ($i=0$). 
At $z=0$, the equivalent current density distribution required to 
create this field is given by the extinction theorem, Eq.~(\ref{E-Field 
Vacuum}), with $\mathbf{r}\in V_1$. Its left hand side vanishes, hence the 
incoming field has to be equated with the remaining surface integrals 
containing two contributions. The first is a convolution of the Green tensor 
with the electric surface current distribution 
$\mathbf{n}(\mathbf{r}')\times\mathbf{B}_0(\mathbf{r}',\omega)$, and the second
is a convolution of the magnetic Green tensor with the magnetic surface current 
distribution $\mathbf{n}(\mathbf{r}')\times\mathbf{E}_0(\mathbf{r}',\omega)$. 
As such, Eq.~(\ref{E-Field Vacuum}) is merely a different form of the source 
quantity representation (\ref{FormalE-Solution}). If we place another medium 
$i=1$ in the bottom half space, the extinction theorem becomes Eq.~(\ref{E-Field 
Medium}) with $\mathbf{r}\in V_0$. Together with the Maxwell boundary 
conditions (\ref{MBC}) one can solve the extinction theorem for the surface 
current densities. Then, Huygens' principle, Eq.~(\ref{E-Field Vacuum}) with 
$\mathbf{r}\in V_0$ and Eq.~(\ref{E-Field Medium}) with $\mathbf{r}\in V_1$ 
has to be used to construct the respective fields ${\mathbf{E}_0}$ and 
$\mathbf{E}_1$. This procedure can be used to derive the reflection 
coefficients of a spatially dispersive medium without having to resort to 
ABC's. For a spherical geometry this has been demonstrated in detail in 
Ref.~\cite{Schmidt2016}.

\subsection{Scattering at a planar halfspace}
\label{subsec:HalfSpace}

We now apply the concepts of Sec.~\ref{subsec:Ex&Huy} to the scattering at a 
spatially dispersive halfspace $i=1$ located at $z<0$. The incoming field is 
impinging onto the surface from the upper halfspace $i=0$, which we assume to 
be free space. We seek to find the reflection coefficients in cylindrical 
coordinates in preparation for Sec.~\ref{sec:SpherePlate} where we consider 
the heat transfer between a plate and a sphere for which it will be necessary 
to transform spherical waves into cylindrical waves. For this reason we expand 
the electromagnetic field as
\begin{gather}
 \begin{split}
\mathbf{E}(\mathbf{r})&=
\sum_n\int\opn{d}\beta\int\opn{d}q\;q\bigg[\alpha_n(\mathbf{q},\beta) 
\mathbf{M}_n(\mathbf{q},\beta,\mathbf{r}) 
+\bar{\beta}_n(\mathbf{q},\beta) 
\mathbf{N}_n(\mathbf{q},\beta,\mathbf{r})
 +\gamma_n(\mathbf{q},\beta) \mathbf{L}(\mathbf{q},\beta,\mathbf{r})\bigg]
\end{split}
\label{MNL(k)-basis}
\end{gather}
in terms of vector cylindrical harmonics $\mathbf{M}_{n}$, $\mathbf{N}_{n}$ and 
$\mathbf{L}_{n}$. In this basis, the bulk Green function (\ref{Bulk-Green}) is 
diagonal and reads
\begin{gather}
\begin{split}
\ten{G}_l(\mathbf{r},\mathbf{r}',\omega)&
=\sum_n \int\limits^{\infty}_0\int\limits^{\infty}_{-\infty}
\opn{d}\beta \opn{d}q\,q 
\bigg\lbrace \frac{\mathbf{M}_{n}(\mathbf{q},\beta,\mathbf{r})
\otimes\mathbf{M}_{-n}(-\mathbf{q},-\beta,\mathbf{r}')}
{D^{(l)}_{\bot}( k,\omega)}\\&
+\frac{\mathbf{N}_{n}(\mathbf{q},\beta,\mathbf{r})
\otimes\mathbf{N}_{-n}(-\mathbf{q},-\beta,\mathbf{r}')}
{D^{(l)}_{\bot}(k,\omega)} 
-\frac{\mathbf{L}_{n}(\mathbf{q},\beta,\mathbf{r})
\otimes\mathbf{L}_{-n}(-\mathbf{ q},-\beta,\mathbf{r}')}
{D^{(l)}_{\parallel}(k,\omega)}\bigg\rbrace\:.
\end{split}\label{CylindricalBulkGreenTensor}
\end{gather}

The surface is in the $(x,y)$ plane at $z=0$. Hence, we can decompose the 
vectors $\mathbf{r}$ and $\mathbf{k}$ into 
$\mathbf{r}=\rho\mathbf{e}_{\rho}+z\mathbf{e}_z$ and 
$\mathbf{k}=\mathbf{q}+\beta\mathbf{e}_z$. The surface integral is to be 
performed over $\opn{d}a(\mathbf{r}')=\rho~\opn{d}\rho~\opn{d}\theta$ with 
$\theta\in[0,2\pi]$ and $\rho\in[0,\infty)$ at $z=0$. 
In order to solve Eqs.~(\ref{E-Field Vacuum}) and (\ref{E-Field Medium}) we 
have to expand the fields in an appropriate basis to make use of the 
convolution theorem. Here we choose the basis 
$\mathbf{X}_n(\mathbf{q},\theta,\rho)$ and 
$\mathbf{e}_z\times\mathbf{X}_n(\mathbf{q},\theta,\rho)$ as vectors in the 
$(x,y)$ plane and ${\bs \chi}_n(\mathbf{q}, \theta, \rho)$ parallel to 
$\mathbf{e}_z$ (App.~\ref{Appendix:BasisVectors}). 
In contrast to the $\mathbf{M}_n$, $\mathbf{N}_n$ and 
$\mathbf{L}_n$ basis, these basis functions only depend on  $\theta,\rho$ which
is convenient when performing the surface integral. The field expansion then 
becomes
\begin{gather}
 \begin{split}
\mathbf{F}_i(\mathbf{r})&=\sum_n\int\opn{d}q\;q
\bigg[A^{(i)}_n(\mathbf{q},z) \mathbf{e}_z
\times\mathbf{X}_{n}(\mathbf{q},\phi,\rho)
+B^{(i)}_n(\mathbf{q},z) 
\mathbf{X}_{n}(\mathbf{q},\phi,\rho)
+C^{(i)}_n(\mathbf{q},z){\bs \chi}_{n}(\mathbf{q},\phi,\rho)\bigg]\;,
\end{split}
\label{fieldExpansionPlate}
\end{gather}
where $A,B,C$ are the expansion coefficients of the electric field and $a,b,c$ 
are those of the magnetic field, respectively. For more details about the 
orthogonality relations and how the bases transform into one another, see 
App.~\ref{Appendix:BasisVectors}. These will be used to transform the 
coefficients $A,B,C$ in Eq.~(\ref{fieldExpansionPlate}) into $\alpha_n$, 
$\bar{\beta}_n$ and $\gamma_n$ necessary for 
Eq.~(\ref{MNL(k)-basis}). 

To this end, we use the MBCs (\ref{MBC}) together with the field expansion 
(\ref{fieldExpansionPlate}) as well as the bulk Green tensor  (\ref{CylindricalBulkGreenTensor})
in Huygens' principle and the extinction 
theorem Eqs.~(\ref{E-Field Vacuum}) and 
(\ref{E-Field Medium}). After performing the surface convolution integral 
we find
\begin{eqnarray}
\mathbf{E}_0(\mathbf{r},\omega)\Theta(z)
&=&\mathbf{E}_{inc}(\mathbf{r},\omega)
+\mathbf{ E}^{(0)}_{G}(\mathbf{r})+\mathbf{E}^{(0)}_{\Gamma}(\mathbf{r}) \nonumber\\
\mathbf{E}_1(\mathbf{r},\omega)\Theta(-z)
&=&-\mathbf{E}^{(1)}_{G}(\mathbf{r})
-\mathbf{ E}^{(1)}_{\Gamma}(\mathbf{r})\:,\label{E-Temp}
\end{eqnarray}
with
\begin{gather}
\mathbf{E}^{(l)}_{G}(\mathbf{r})=i\omega\sum_n\int\opn{d}q\;q 
\bigg\lbrace 
b^{(0)}_{n}(\mathbf{q},0)Z^{(l)}_s(z,0)
\mathbf{e}_z\times\mathbf{X}_{n}(\mathbf{q}, \phi,\rho) \nonumber \\ 
-a^{(0)}_{n}(\mathbf{q},0)Z^{(l)}_q(z,0)\mathbf{X}_{n}(\mathbf{q},\phi,\rho) 
+a^{(0)}_{n}(\mathbf{q},0)Z^{(l)}_z(z,0){\bs \chi}_n(\mathbf{q},\phi,\rho)
\bigg\rbrace
\end{gather}
and
\begin{gather}
\mathbf{E}^{(l)}_{\Gamma}(\mathbf{r})=-i\sum_n(-1)^n\int\opn{d}
q\;q 
\bigg\lbrace
A^{(0)}_n(\mathbf{q},0)\zeta^{(l)}_s(z,0)\mathbf{e}_z
\times\mathbf{X}_{-n}(-\mathbf{q},\phi,\rho) \nonumber\\
+B^{(0)}_n(\mathbf{q},0)\big[q\zeta^{(l)}_q(z,0)
{\bs \chi}_{-n}(-\mathbf{q},\phi,\rho) 
+\zeta^{(l)}_s(z,0)\mathbf{X}_{-n}(-\mathbf{q},\phi,\rho)\big] \bigg\rbrace\; .
\end{gather}

In order to simplify the $\beta$ integration, which has its origin in the Green tensor
Eq.~(\ref{CylindricalBulkGreenTensor}), we defined the surface impedances
\begin{gather}
 \begin{split}
Z^{(l)}_s(z,z')
&=\int\limits^{\infty}_{-\infty}\frac{\opn{d}\beta}{2\pi} 
\frac{e^{i\beta(z+z')}}{D^{(l)}_{\bot}( q,\beta,\omega)},\\
Z^{(l)}_q(z,z')
&=\int\limits^{\infty}_{-\infty}\frac{\opn{d}\beta}{2\pi}
\frac{e^{i\beta(z+z')}}{\beta^2+q^2} 
\Big[\frac{\beta^2}{D^{(l)}_{\bot}(q,\beta,\omega)}
-\frac{q^2}{D^{(l)}_{\parallel}(q,\beta,\omega)}\Big],\\
Z^{(l)}_z(z,z')
&=\int\limits^{\infty}_{-\infty}\frac{\opn{d}\beta}{2\pi}
\frac{q\beta e^{i\beta(z+z')}}{\beta^2+q^2} 
\Big[\frac{1}{D^{(l)}_{\bot}(q,\beta,\omega)} 
+\frac{1}{D^{(l)}_{\parallel}(q,\beta,\omega)}\Big],
\end{split}
\end{gather}
as well as
\begin{gather}
\zeta^{(l)}_s(z,z')=\int\limits^{\infty}_{-\infty}\frac{\opn{d}\beta}{2\pi}
\frac{\beta e^{-i\beta(z-z')}}{D^{(l)}_{\bot}( q,\beta,\omega)}, \nonumber\\
\zeta^{(l)}_q(z,z')=\int\limits^{\infty}_{-\infty}\frac{\opn{d}\beta}{2\pi}
\frac{ e^{-i\beta(z-z')}}{D^{(l)}_{\bot}( q,\beta,\omega)}.
\end{gather}
In the local limit, i.e. when spatial dispersion can be disregarded, these 
integrals are known and have been calculated previously \cite{Chew1995}. Note
that only $Z^{(l)}_q$ and $Z^{(l)}_z$ contain the longitudinal dispersion 
relation $D^{(l)}_{\parallel}$. We also note in passing that $Z_q$ and $Z_s$ 
are, up to an irrelevant prefactor, the well-known Fuchs-Kliewer impedances 
\cite{Kliewer1968, Kliewer1971, Fisher1975}. 

When expanding the left-hand side of Eq.~(\ref{E-Temp}) as well as the incoming field
we find a conditional equation for the field amplitudes corresponding to Eq.~(\ref{E-Field Vacuum}), i.e.
\begin{eqnarray}
A^{(0)}_n(\mathbf{q},z)\Theta(z)
&=&\big[A^{inc}_n(\mathbf{q},z)+i\omega 
b^{(0)}_{n}(\mathbf{q},0)Z^{(0)}_s(z,0) +iA_n(\mathbf{q},0)\zeta^{(0)}_s(z,0) 
\big] \nonumber\\
B^{(0)}_n(\mathbf{q},z)\Theta(z)
&=&\big[B^{inc}_n(\mathbf{q},z)-i\omega 
a^{(0)}_{n}(\mathbf{q},0)Z^{(0)}_q(z,0)+iB_n(\mathbf{q},0)\zeta^{(0)}_s(z,0)\big] \nonumber\\
C^{(0)}_n(\mathbf{q},z)\Theta(z)
&=&\big[c^{inc}_n(\mathbf{q},z)+i\omega 
a^{(0)}_{n}(\mathbf{q},0)Z^{(0)}_z(z,0) -iB_n(\mathbf{q},0)q\zeta^{(0)}_q(z,0)\big].
\label{Extinction theorem (Vacuum)}
\end{eqnarray}
The extinction theorem is obtained by taking the limit $z\nearrow 0^-$ which is 
equivalent to demanding $\mathbf{E}^{(0)}_{inc}+\mathbf{E}^{(0)}_{scat}=0$ at 
the boundary. 
Together with the second part of the extinction theorem, that is, by taking the 
limit $z\searrow 0^{+}$ inside the medium Eq.~(\ref{E-Field Medium}), viz.
\begin{gather}
 0=\omega b^{(0)}_{n}(\mathbf{q},0)
 Z^{(1)}_s(0^{+},0)+A^{(0)}_n(\mathbf{q},0)\zeta^{(1)}_s(0^{+},0), \nonumber\\
  0=-\omega a^{(0)}_{n}(\mathbf{q},0)
 Z^{(1)}_q(0^{+},0)+B^{(0)}_n(\mathbf{q},0)\zeta^{(1)}_s(0^{+},0) \nonumber\\
  0=\omega a^{(0)}_{n}(\mathbf{q},0)
 Z^{(1)}_z(0^{+},0) -B^{(0)}_n(\mathbf{q},0)q\zeta^{(1)}_q(0^{+},0),
\end{gather}
one is able to derive the field amplitudes. Thus we find as an intermediate result,
\begin{gather}
A^{(0)}_n(\mathbf{q},0)= 
-\frac{iA^{inc}_n(\mathbf{q},0)Z^{(1)}_s(0^{+},0)}
{\zeta^{(1)}_s(0^{+},0)Z^{(0)}_s(0^{-},0)-Z^{(1)}_s(0^{+},0)
\zeta^{(0)}_s(0^{-}, 0)}, \nonumber \\
B^{(0)}_n(\mathbf{q},0)= 
-i\frac{B^{inc}_n(\mathbf{q},0)Z^{(1)}_q(0^{+},0)}
{\zeta^{(1)}_s(0^{+},0)Z^{(0)}_q(0^{-},0)-\zeta^{(0)}_s(0^{-},0)
Z^{(1)}_q(0^{+}, 0)}.
\end{gather}

In the free-space region there are no additional bulk modes. For this reason 
the $C^{(0)}_n$ are not required as they are linearly dependent on 
$B^{(0)}_n$, see App.~\ref{Appendix:BasisVectors}. Next, we use Huygens' 
principle, Eq.~(\ref{Extinction theorem (Vacuum)}), with $z>0$ and decompose 
the field in the free-space region into an incoming $\mathbf{E}^{(0)}_{inc}$
and scattering $\mathbf{ E}^{(0)}_{scat}$ part. In the scattered field 
$\mathbf{E}^{(0)}_{scat}$, the components $A^{scat}_n(\mathbf{q},z)= 
A^{inc}_n(\mathbf{q},0)R_s(\mathbf{q},z)$ and 
$B^{scat}_n(\mathbf{q},z)=B^{inc}_n(\mathbf{q},0)R_q(\mathbf{q},z)$ are 
proportional to their respective reflection coefficients. We thus obtain the 
reflection coefficients as
\begin{gather}
R_s(\mathbf{q},z)= 
-\frac{\zeta^{(1)}_s(0^{+},0)Z^{(0)}_s(z,0)-Z^{(1)}_s(0^{+},0)
\zeta^{(0)}_s(z,0)}{\zeta^{(1)}_s(0^{+},0)Z^{(0)}_s(0^{-},0)-Z^{(1)}_s(0^{+},0)
\zeta^{(0)}_s(0^{-},0)} \nonumber\\
R_q(\mathbf{q},z)= 
-\frac{\zeta^{(1)}_s(0^{+},0)Z^{(0)}_q(z,0)-Z^{(1)}_q(0^{+},0)
\zeta^{(0)}_s(z,0)}{\zeta^{(1)}_s(0^{+},0)Z^{(0)}_q(0^{-},0)
-\zeta^{(0)}_s(0^{-},0)Z^{(1)}_q(0^{+},0)}\:,\label{reflxcoef}
\end{gather}
which still depend on $z$. The impedances describing the free-space region can 
be evaluated analytically to
\begin{equation}
R_s(\mathbf{q},z)=r_{s}(q_0)e^{i\beta^{+}_0z},\:
R_q(\mathbf{q},z)=-r_p(q_0)e^{i\beta_0z}
\label{ReflexPlateTemp}
\end{equation}
with
\begin{gather}
r_{s}(q_0)=-\frac{\zeta^{(1)}_s(0^{+},0)+\beta^{+}_0Z^{(1)}_s(0^{+},0)}
{\zeta^{(1)}_s(0^{+},0)-\beta^{+}_0Z^{(1)}_s(0^{+},0)}, \nonumber\\
r_p(q_0)=\frac{\beta_0\zeta^{(1)}_s(0^{+},0)+k^2_0Z^{(1)}_q(0^{+},0)}
{\beta_0\zeta^{(1)}_s(0^{+},0)-k^2_0Z^{(1)}_q(0^{+},0)}.
 \label{ReflexPlateFinal}
\end{gather}
In the case of a local medium $i=1$, the usual Fresnel reflection coefficients 
are retrieved. The minus sign in Eq.~(\ref{ReflexPlateTemp}) has its origin in 
the definition basis functions (App.~\ref{Appendix:Basetransformations}).

\subsection{Numerical results of the reflection coefficients}

The reflection coefficients in Eq.~(\ref{ReflexPlateFinal}) depend on the 
impedances $\zeta^{(1)}_s$, $Z^{(1)}_s$ and $Z^{(1)}_q$. All of them are 
integrals over the dispersion relations $D^{(1)}_{\bot}$ and $D^{(1)}_{\|}$ for 
which the longitudinal and transversal dielectric functions have to be 
specified. Calculating these integrals analytical is feasible, but only for 
algebraic dielectric functions. One such function has been introduced in
Eq.~(\ref{Epsilon_ZnSe}). The longitudinal and transversal dielectric response 
function differ only in their spatial dispersion parameter 
$D_{\|}=(1+\delta)D_{\bot}$.

\begin{figure}[ht]
\centering
\subfloat{{\includegraphics[width=8.3cm]{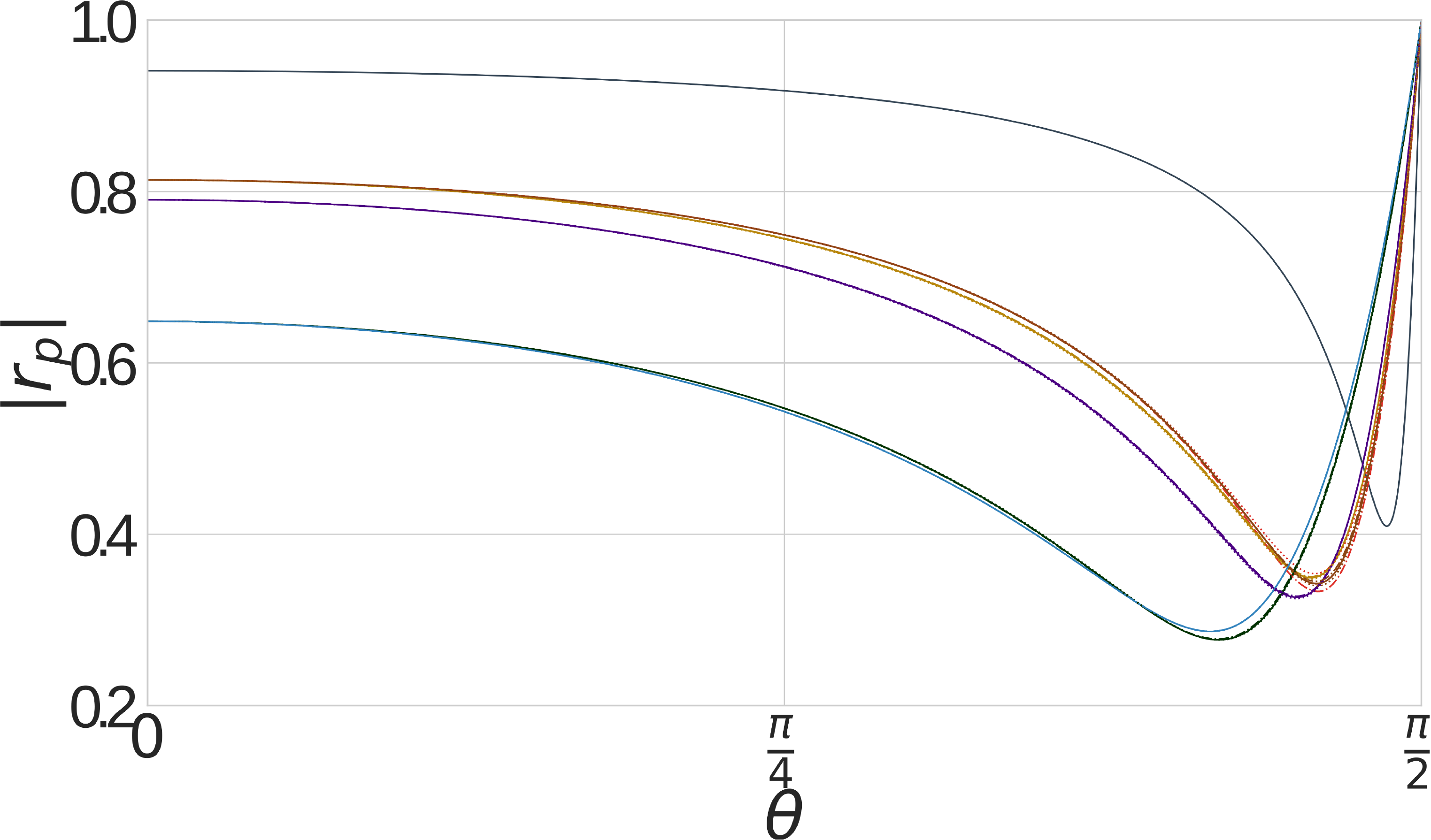}}}%
\subfloat{{\includegraphics[width=8.3cm]{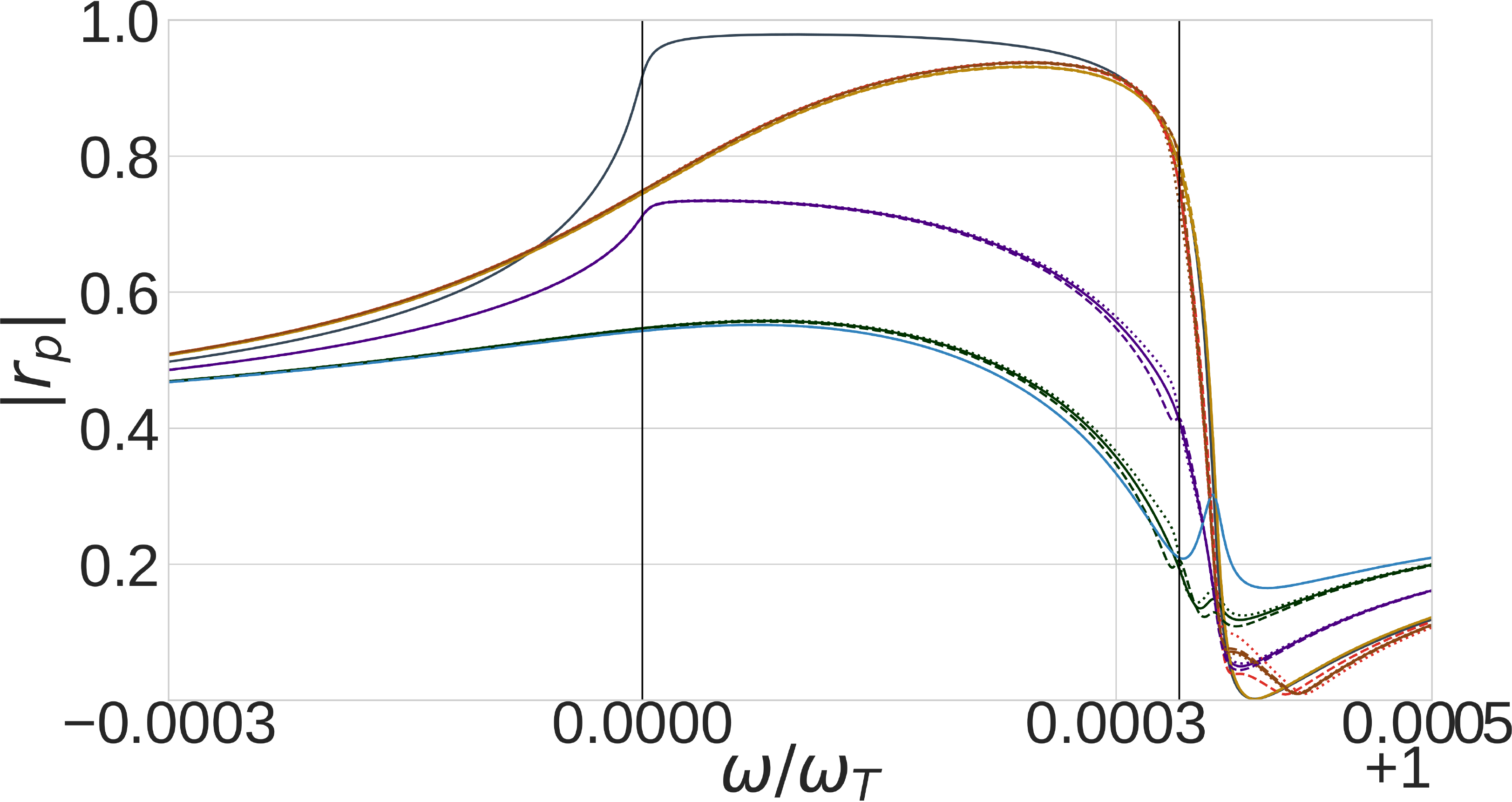}}}%

\caption{\label{fig:reflex1} Absolute value of the reflection coefficient of a 
planar ZnSe surface. Left: reflection coefficient as function of incident angle 
$\theta$ at $\omega=\omega_T$. Right: reflection coefficient as function of 
frequency at $\theta=\pi/4$ with $\delta=0$ (solid 
lines), $\delta=0.5$ (doted) and $\delta=-0.5$ (dashed) and the color coding: 
Local (grey), HuyEx (red), Ting \textsl{et al.}  (dark gold), Pekar (green),  
Agarwal \textsl{et al. } (indigo), Fuchs-Kliewer (brown), Rimbey-Mahan (ice 
blue).}
 \end{figure}

A material whose electromagnetic response can be cast into such an algebraic 
form which is used to investigate the influence of spatial dispersion is ZnSe. 
For this material, the reflection coefficients are plotted in 
Fig.~\ref{fig:reflex1} as a function of the incident angle $\theta$ as well as 
the frequency $\omega$ for different values of $\delta$. 
We compare the results derived from Huygens' principle and the extinction 
theorem to the local case and ABCs from various sources:  Agarwal 
\textit{et al.} 
\cite{Agarwal1975,Agarwal1971,Agarwal1971b,Agarwal1972,Agarwal1973}, Ting 
\textit{et al.} \cite{Ting1975}, Fuchs-Kliewer 
\cite{Kliewer1968,Kliewer1971,Fisher1975}, Rimbey-Mahan 
\cite{Rimbey1974,Rimbey1975,Johnson1976,Rimbey1977,Rimbey1978} and Pekar 
\cite{Pekar1958a,Pekar1958b,Pekar1958c,Pekar1959}.
For a general derivation of these ABCs, taking the longitudinal-transversal 
splitting into account, the interested reader is referred to a recent 
work by Churchill and  Philbin \cite{Churchill2016}.

It is evident that spatial dispersion significantly  reduces the reflectivity 
for most of the angular spectrum at the transition frequency $\omega_T$. Only 
when the incident field arrives at a shallow angle, is the reflectivity 
increased. The Pekar and Rimbey-Mahan ABC's  predict the lowest reflectivity 
for frequencies smaller than $\omega_L$. The Fuchs-Kliewer and Ting \textsl{et 
al.} boundary conditions resemble the results obtained from the Huygens'
principle and the extinction theorem. In fact, the Fuchs-Kliewer boundary 
condition matches quite well the HuyEx results, although in a narrow frequency 
and angular interval, the transverse-longitudinal splitting $\delta\neq0$ is 
less pronounced.

\section{Thermal emissivity of an isolated spatially dispersive sphere}
\label{sec:sphere-emission}

Now we turn to the thermal emissivity of a spatially dispersive sphere.
The spectral energy flux mediated by the electromagnetic field from a sphere
of radius $R$ at temperature $T_{sp}$ into free space through a surface 
$\mathcal{A}$ is given by
\begin{equation}
S(\omega,T_{sp})=\mu^{-1}_0\opn{Re} \int\limits_{\mathcal{A}}
\opn{d}\mathbf{a}(\mathbf{r})\cdot
\big\langle{\bs E}(\mathbf{r},\omega)\times{\bs B}^\ast(\mathbf{r},\omega) 
\big\rangle_{T_{sp}}\label{SFlux}
\end{equation}
where $\langle ... \rangle_{T_{sp}}$ denotes the thermal average over the 
spherical volume. The electromagnetic field $\mathbf{E}(\mathbf{r},\omega)$ is 
a solution of the Helmholtz equation. If we use the source quantity 
representation (\ref{FormalE-Solution}), we can write the expectation value as
\begin{gather}
\big\langle \mathbf{E}(\mathbf{r},\omega)\times 
\mathbf{B}^\ast(\mathbf{r}',\omega)\big\rangle_T
=-\mu^3_0\omega^2\int\limits_{V_{\opn{sp}}}\opn{d} ^3s
\int\limits_{V_{\opn{sp}}}\opn{d}^3s' 
\opn{O_{c}}\big\lbrace \ten{G}(\mathbf{r},\mathbf{ s})\cdot \big\langle 
\mathbf{j}(\mathbf{s},\omega)\otimes \mathbf{j}^\ast(\mathbf{s}',\omega) 
\big\rangle_T\cdot\mathbf{\Gamma}^{\dagger}(\mathbf{r}',\mathbf{s}') 
\big\rbrace\;,
\label{ExB}
\end{gather}
where ${\bs \Gamma}(\mathbf{r},\mathbf{r'},\omega) 
=-(i\mu_0\omega)^{-1}\nabla\times \ten{G}(\mathbf{r},\mathbf{r}',\omega)$ is 
again the magnetic dyadic Green function. We also introduced the outer vector 
product $\opn{O_{c}}\lbrace\cdots\rbrace$, which is defined as the curl between 
the outermost left and right vectors 
$\opn{O_{c}}\lbrace\mathbf{A}\otimes\mathbf{B}\rbrace
=\mathbf{A}\times\mathbf{B}$.
Note, that the vectors $\mathbf{r}$ and $\mathbf{r}'$ remain outside 
$V_{\opn{sp}}$.

In thermal equilibrium and within linear response theory, we can apply the 
fluctuation-dissipation theorem to compute the thermal expectation value of the 
current density as
\begin{equation}
\big\langle \mathbf{j}(\mathbf{r},\omega)\otimes 
\mathbf{j}^\ast(\mathbf{r}',\omega) \big\rangle_T
=2\omega\epsilon_0\opn{Im}\ten{\epsilon}(\mathbf{r},\mathbf{r}', 
\omega)\Theta(\omega,T)\:.
\label{FDT} 
\end{equation}
Here, $\Theta(\omega,T)$ refers to the thermal distribution function 
$\Theta(\omega,T)=\hbar \omega /(e^{\hbar\omega/k_bT}-1)$. One can then
derive the spectral heat transfer rate $S(\omega,T_{sp})$ in terms of the 
dyadic Greens function applying some algebra (for details, see 
App.~\ref{Appendix:ExB}) as 
\begin{gather}
S(\omega,T_A)=\mu_0\omega\Theta(\omega,T)\opn{Im}
\int\limits_{\mathcal{A}}\opn{d}\mathbf{a}(\mathbf{r})\cdot 
\int\limits_{\partial V_{\opn{sp}}}\opn{d}^2 s\;\opn{O_{p}} 
\Big\lbrace \big[\mathbf{n}(\mathbf{s})\times
\ten{G}^{T}(\mathbf{r},\mathbf{s})\big]\times
\big[\mathbf{\Gamma}^\ast(\mathbf{r}',\mathbf{s})
\times\overleftarrow{\nabla}_s\big] \nonumber \\
-\mathbf{n}(\mathbf{s})\times\big[\nabla_s\times
\ten{G}^{T}(\mathbf{r},\mathbf{s})\big]\times
\mathbf{\Gamma}^\ast(\mathbf{r},\mathbf{s})\Big\rbrace\;.
 \label{Poitning TheoremNLSphere}
\end{gather}
Here we introduced the notation for the outer scalar product 
$\opn{O_{p}}\lbrace\cdots\rbrace$ which is defined as the scalar product 
between the outermost left and right vectors 
$\opn{O_{p}}\lbrace\mathbf{A}\otimes\mathbf{B}\rbrace
=\mathbf{A}\cdot\mathbf{B}$.

Equation~(\ref{Poitning TheoremNLSphere}) contains two surface integrals, one 
extends over the surface $\partial V_{sp}$ encapsulating the source region, the 
other covers the surface $\mathcal{A}$ through which the energy flux has to be 
computed. The latter is taken to be another spherical surface with radius 
$\rho>R$. We decompose the Green tensor into a freely propagating and a 
scattering part according to Eq.~(\ref{LinarGreen}) which can be found in the 
literature, see e.g. Ref.~\cite{Li1994}. In contrast to the local case, we use 
the reflection coefficients that take spatial dispersion into account.
Performing the surface integrals and making use of the Wronskian 
of the spherical Hankel functions, we obtain a result that is independent 
of $\rho$, 
\begin{gather}
S(\omega,T)=(k_0R)^2\Theta(\omega,T)e(\omega)\\
=-2k^2_0R^2\Theta(\omega,T)\sum\limits^{\infty}_{n=1}(2n+1) 
\big(\opn{Im}[\tilde{R}^{1,M}_n\tilde{R}^{2,M\ast}_n]
+\opn{Im}[\tilde{R}^{1,N}_n\tilde{R}^{2,N\ast}_n]\big). \nonumber \\
\label{SpEmission}
\end{gather}
where $e(\omega)$ denotes the spectral emissivity.
This can be further simplified by applying the Wronskian once again to obtain
\begin{equation}
\opn{Im}[\tilde{R}^{1,X}_n\tilde{R}^{2,X\ast}_n]
=\frac{\opn{Re}\big[r^{X}_n\big]}{k^2_0R^2}
-\frac{|r^{X}_n|^2}{k^2_0R^2}
\end{equation}
with $X\in\big\lbrace \mathbf{M},\mathbf{N}\big\rbrace$. This result corresponds
to findings previously obtained in the local case \cite{Kattawar1970}.

\begin{figure}[ht]
\centering
\includegraphics[width=12.5cm]{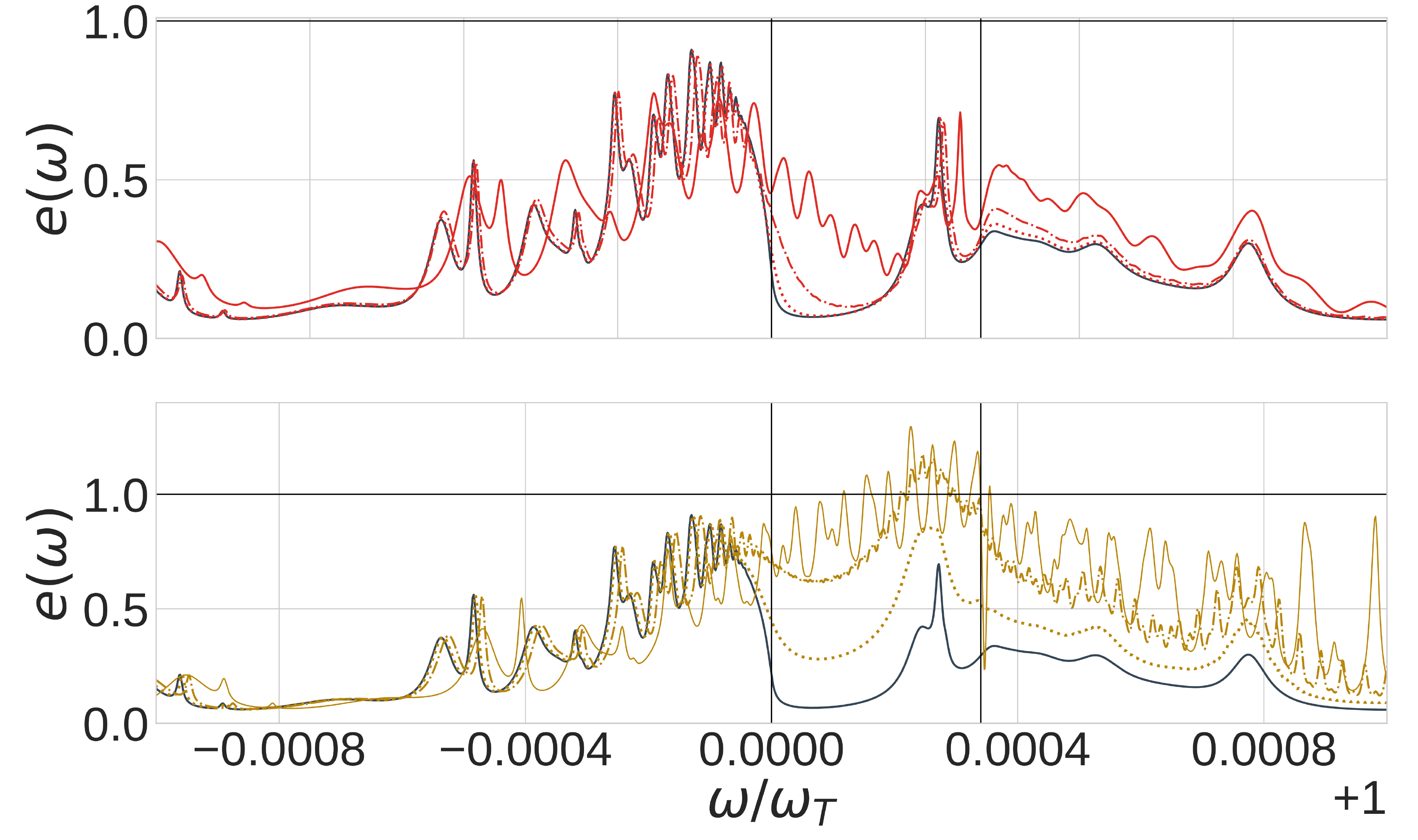}
\caption{\label{fig:emissivity} Depicts a comparison of local (black) versus the nonlocal
spectral emissivity for a ZnSe sphere with $R=220$nm for different dispersion parameters:
$D=D_{ZnSe}$ (solid), $D=D_{ZnSe}/10$ (dashed) and $D=D_{ZnSe}/100$ (dotted). The local 
case is compared to HuyEx (red, top) and Pekar (gold, bottom).}
\end{figure}

Figure~\ref{fig:emissivity} shows the emissivity of ZnSe as a function of frequency
for a sphere of radius  $R=220$nm resulting in a size parameter $k_0R=3.12>1$. 

There is a noticeable difference between the local and nonlocal results. The 
overall emissivity is increased, and the spectral features are shifted towards 
higher frequencies. This is more pronounced for the Pekar ABC than for the HuyEx
results. In both cases, the emissivity converges towards the local results 
with decreasing dispersion parameter $D=D_{ZnSe}$.
The Pekar ABC predict plenty of spectral features, and the emissivity is 
highest within the stop band, exceeding $e(\omega)=1$. In contrast, the HuyEx
results predict the maximum emissivity below the stop band and a lower 
emissivity. 

We also notice that there are frequencies for which the emissivity is enhanced 
beyond the boundary of a perfect Planckian emitter, i.e. $e(\omega)>1$.  That 
is a well-understood phenomenon \cite{Book:Bohren2002, Biehs2016} which becomes 
more pronounced with decreasing radius.  
A perfect black body is one that absorbs all radiation incident on it.  However,
this standard definition has a geometric component to it. Due to the wave 
nature of light, this definition is only sensible when the size of the object 
is 
much larger than the wavelength.  As soon as the size of an object becomes 
comparable to the wavelength, its geometric cross-section becomes smaller than 
the absorption/emission cross-section which implies that the emissivity 
can exceed $e(\omega)=1$.

\section{Radiation driven near-field heat transfer between a sphere and a 
plate}
\label{sec:SpherePlate}

\begin{figure}[ht]
\centering
\includegraphics[width=8.5cm]{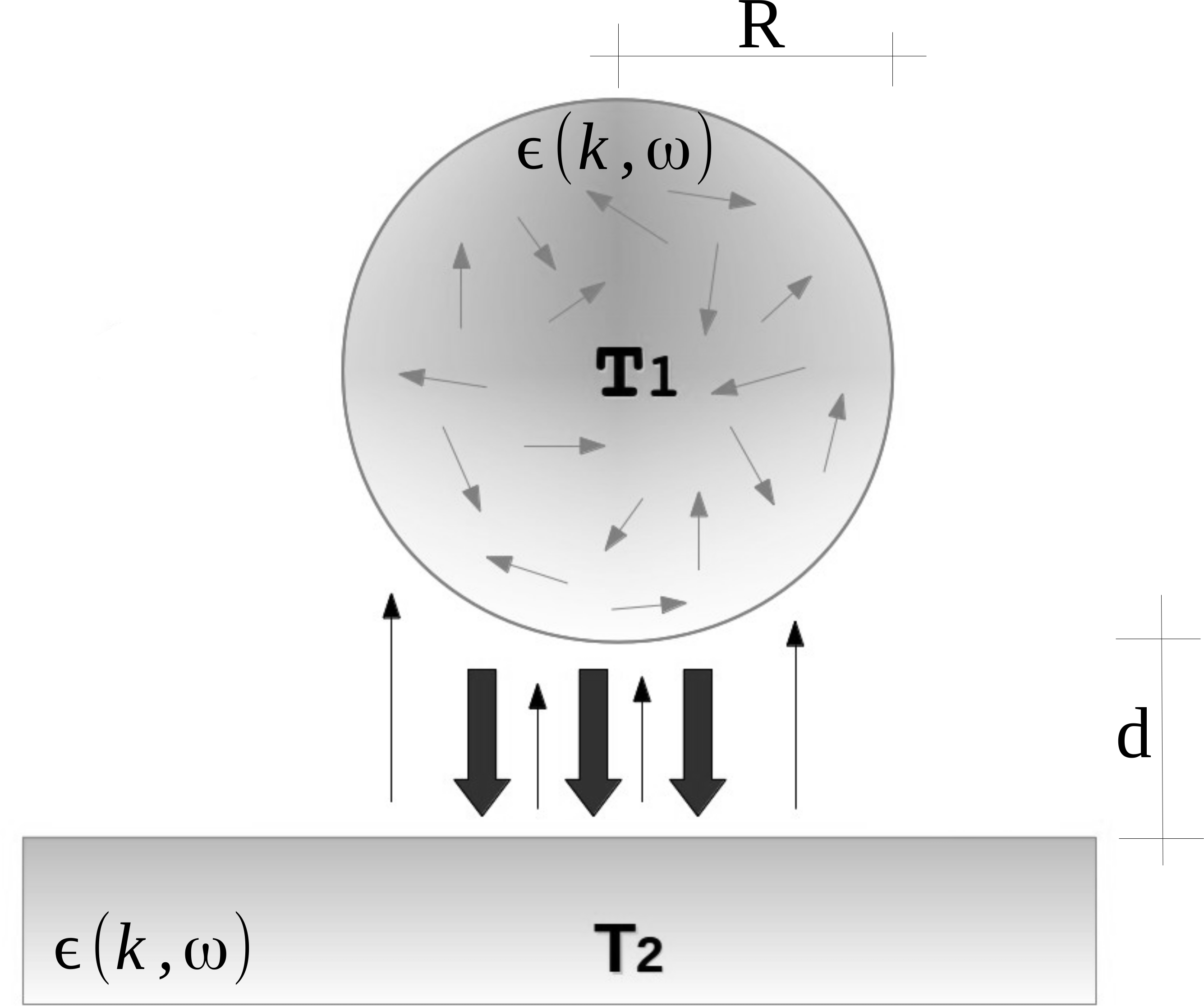}
\caption{\label{fig:sphere-plate} Schematic set up with a dielectric sphere of 
Radius $R$ at a distance $d$ above a dielectric half-space. Both are held at 
constant temperatures $T_{sp}$ and $T_{pl}$, respectively.}
\end{figure}

We now turn to the problem of combining the emission from a sphere with that of 
a planar halfspace and compute the radiative heat transfer between them.
We consider a plate $pl$ in the $(x,y)$ plane that occupies the entire lower 
half space $z\leq 0$. At a distance $d$  a sphere $sp$ with radius $R$ is 
embedded into free space. Its centre of mass is located  at 
$z=d+R$, see Fig.~\ref{fig:sphere-plate}. The sphere is held at a temperature 
$T_{sp}$ and the plate at temperature $T_{pl}$, respectively. We assume the 
temperatures to be constant, and the energy flow is mediated only by 
electromagnetic fields. Then both objects emit and absorb thermal radiation at 
a constant rate. It is sufficient to consider the effective energy transfer in 
one direction only, to establish the net heat transfer, we will apply the 
reciprocity condition.

We begin this section by briefly reviewing the approach to heat transfer 
between dielectric bodies \cite{Otey2011} and introduce the necessary 
modifications to include spatial dispersion. In the previous sections, we
have presented the derivation of the plate reflection coefficients and the 
emissivity of an isolated sphere. In each derivation, we required basis vectors 
of their particular geometry, i.e. either vector spherical or cylindrical 
harmonics.  For the joint geometry of plate and sphere, it will be necessary
to interconvert between the bases.

Let ${\bs \Psi}^{(\pm)}_{\alpha}(k,\mathbf{ r})$ represent the set of 
vector spherical harmonics, where  $\alpha=\lbrace \mathbf{ P}, n,m\rbrace$ is a 
multi-index containing the polarisation 
$\mathbf{ P}\in\lbrace \mathbf{ M},\mathbf{ N}\rbrace$ and the indices
$n\in\mathbb{N}$ as well as  $m\in \big[-n,n\big]$. The origin of the vector 
spherical harmonics is at the centre of the sphere. The superscript $(\pm)$
denotes an outgoing $(+)$ or an incoming wave $(-)$, relative to the 
origin of the reference frame, and requires one to replace the Bessel 
function by the Hankel function of the first kind 
(superscript $(+)\rightarrow(1)$) and the Hankel function of the second kind 
(superscript $(-)\rightarrow(2)$), respectively. In this notation the electric 
field outside an isolated, radiating sphere becomes
\begin{gather}
\mathbf{ E}_{sp}(\mathbf{ r},\omega)=\sum\limits_{\alpha}\mathcal{A}^{0}_{\alpha+}
{\bs\Psi}^{(+)}_{\alpha}(k_0,\mathbf{ r}).
\label{sphere-field expansion(emission)}
\end{gather}
It describes waves propagating away from the sphere with amplitude 
$\mathcal{A}^{0}_{\alpha+}$.  In this notation, Eqs.~(\ref{SFlux}) and 
(\ref{SpEmission}) become
\begin{gather}
\big\langle \mathbf{ E}_{sp}(\mathbf{ r},\omega)\times 
\mathbf{ B}_{sp}^{*}(\mathbf{ r}',\omega)\big\rangle_T
=\mathcal{N}(\omega,T) e^{}_{\alpha}(\omega)
{\bs\Psi}^{(+)}_{\alpha}(k_0,\mathbf{ r})\times\big[\nabla'\times 
{\bs\Psi}^{(+)*}_{\alpha,}(k_0,\mathbf{ r}')\big] \nonumber \\
=-i\omega\big\langle\mathcal{A}^{0}_{\alpha^{}_+}
\mathcal{A}^{0*}_{\alpha^{}_+}\big\rangle_T
{\bs\Psi}^{(+)}_{\alpha}(k_0,\mathbf{ r})\times\big[\nabla'\times 
{\bs\Psi}^{(+)*}_{\alpha}(k_0,\mathbf{ r}')\big],
\label{ExB-Spherical-Coord}
\end{gather}
where we introduced the variables
\begin{gather}
e^{}_{\alpha}(\omega)=2ik_0\opn{Im}[\tilde{R}^{1,P}_n\tilde{R}^{2,P*}_n]
\big(\delta_{\mathbf{ P},\mathbf{ N}}+\delta_{\mathbf{ P},\mathbf{ M}}\big),\nonumber \\
\mathcal{N}(\omega, T)=\mu_0k^2_0R^2\Theta(\omega,T).
\end{gather}

For the plate reference system we already introduced the vector cylindrical 
harmonics, here denoted by  ${\bs\phi}^{(\pm)}_{\beta}(\mathbf{ k},\mathbf{ r})$. The 
multi-index $\beta\in\lbrace \mathbf{ P}, n\rbrace$ contains the polarisation 
$\mathbf{ M}, \mathbf{ N}$ and $n\in\mathbb{N}$. The point of origin is on the plate 
surface closest to the sphere, i.e. the direction of the $z$-axis points to the 
centre of the sphere. The superscript $\pm$ refers to outgoing or incoming 
waves, relative to the origin of the plate reference frame.

While Eqs.~(\ref{sphere-field expansion(emission)}) 
and (\ref{ExB-Spherical-Coord}) describe the field emitted by an isolated 
sphere, we require the knowledge of the total emitted field including multiple 
reflections between the sphere and the plate. The total electromagnetic 
field emitted, by the sphere, and in cylindrical coordinates with the origin at 
the plate interface, shall hereby be denoted with
\begin{gather}
\mathbf{ E}^{tot}_{sp}(\mathbf{ r},\omega) =\sum\limits_{\beta}
\mathcal{B}^{}_{\beta}{\bs\phi}^{(+)}_{\beta}(\mathbf{ k}_0,\mathbf{ r}).
\end{gather}
The thermal expectation value of the vector product of the electric and 
magnetic field becomes
\begin{gather}
\big\langle \mathbf{ E}_{}(\mathbf{ r},\omega)\times 
\mathbf{ B}_{}^{*}(\mathbf{ r}',\omega)\big\rangle_T 
=-i\omega\sum\limits_{\beta',\beta}\big\langle 
\mathcal{B}^{}_{\beta}\mathcal{B}^{}_{\beta'}\big\rangle_T
{\bs\phi}^{(+)}_{\beta}(\mathbf{ k}_0,\mathbf{ r})\times
\big[\nabla\times{\bs\phi}^{(+)*}_{\beta'}(\mathbf{ k}_0,\mathbf{ r}')\big]\;.
\label{ExBCylinderCoord}
\end{gather}

Next, we relate the amplitudes $\mathcal{A}^{0}_{\alpha^{}_+}$ of a wave 
emitted by an isolated sphere to its total amplitude $\mathcal{B}^{}_{\beta}$ in 
the presence of a plate. The conversion is described by an operator 
$\mathbb{O}^{}_{\beta\alpha^{}_+}$ as
\begin{gather}
\mathcal{B}^{}_{\beta}= 
\sum\limits_{\alpha}\mathbb{O}^{}_{\beta\alpha^{}_+}
\mathcal{A}^{0}_{\alpha^{}_+}. 
\label{Operator-O}
\end{gather}
To determine this operator we consider a mode $\Pi\in\lbrace\mathbf{ M}^{}_{nm}, 
\mathbf{ N}^{}_{nm}\rbrace$ of spherical waves, emanating from the sphere 
${\bs\Psi}^{(+)}_{\alpha}$ and a corresponding mode emanating from the plate 
${\bs\phi}^{(\pm)}_{\beta}$. The resulting electric field $\mathbf{ E}^{\Pi}$ in the 
vacuum region is then given as a superposition
\begin{gather}
\mathbf{ E}^{\Pi}(\mathbf{ r},\omega)
=\sum\limits_{\alpha}\mathcal{A}^{\Pi}_{\alpha^{}_+}
{\bs\Psi}^{(+)}_{\alpha}(k_0,\mathbf{ r})
+\sum\limits_{\beta}\mathcal{B}^{\Pi}_{\beta^{}_+}
{\bs\phi}^{(+)}_{\beta}(\mathbf{ k}_0,\mathbf{ r}).
\end{gather}

It is useful to introduce another operator $\Lambda^{}_{yx}$ that transforms 
vector spherical into cylindrical waves or vice versa \cite{Han2008}. 
An incoming spherical wave is partially reflected at the plate, which is the 
only source of cylindrical waves. Hence, we can write
\begin{gather}
\mathcal{B}^{\Pi}_{\beta^{}_+}=R^{}_{\beta+\beta'-}
\Lambda^{}_{\beta_-',\alpha^{}_+}\mathcal{A}^{\Pi}_{\alpha^{}_+},
\label{B-Beta}
\end{gather}
where $R^{}_{\beta+\beta'-}$ are the reflection coefficients in cylindrical 
coordinates of the plate, which we derived in Sec.~\ref{sec:Ex&Huy}. An
outgoing spherical wave with amplitude $\mathcal{A}^{\Pi}_{\alpha_+}$ can be 
generated by an emission $\mathcal{A}^{0}_{\alpha^{}_+}$ or by
reflection of an impinging cylindrical wave at the surface of the sphere. Thus 
we find
\begin{gather}
\mathcal{A}^{\Pi}_{\alpha^{}_+}=
\mathcal{A}^{0}_{\alpha^{}_+}\delta^{}_{\alpha^{}_+\Pi}
+r^{}_{\alpha^{}_+\alpha_-'}\Lambda^{}_{\alpha_-',\beta^{}_+}
\mathcal{B}^{\Pi}_{\beta^{}_+}.
\end{gather}
Combined with Eq.~(\ref{B-Beta}) we obtain a conditional equation for 
the total amplitude $A^{\Pi}_{\alpha_+'}$
\begin{gather}
\Big[\delta^{}_{\alpha^{}_+,\alpha_+'}-T^{}_{\alpha^{}_+,\alpha_+'}\Big]
\mathcal{A}^{\Pi}_{\alpha_+'}=
\mathcal{A}^{0}_{\alpha_+'}\delta^{}_{\alpha_+'\Pi }\;,
\label{Conditinal equation total Amplitude}
\end{gather}
with
\begin{gather}
T^{}_{\alpha^{}_+,\alpha_+'}=r^{}_{\alpha_+\tilde{\alpha}_-}
\Lambda^{}_{\tilde{\alpha}^{}_-,\beta_+'}
R^{}_{\beta'_+\beta_-}\Lambda^{}_{
\beta^{}_-,\alpha_+'}.
\end{gather}
Once the amplitudes $\mathcal{A}^{\Pi}_{\alpha^{}_+}$ are
determined from Eq.~(\ref{Conditinal equation total Amplitude}), one is able 
to derive an expression for the conversion operator 
\begin{gather}  
\mathbb{O}_{\beta\alpha+}=\tau_{\beta^{}_+\beta_-'}
\Lambda^{}_{\beta_-',\alpha_+'}
\Big[\delta^{}_{\alpha^{}_-,\alpha_+'}-T^{}_{\alpha^{}_+,\alpha_+'} 
\Big]^{-1}\;.
\end{gather}
We are only interested in the radiation absorbed by the half space, which is 
represented by the transmission coefficients $\tau_{\beta\beta'}$.
However, $\tau_{\beta\beta'}$ is a diagonal matrix, and only the absolute
square will contribute. Hence, we can replace $|\tau_{\beta\beta'}|^2$
with $1-|R_{\beta,\beta'}|^2$. Together with  Eq.~(\ref{Conditinal equation 
total Amplitude}), we can write the Poynting vector as a function of the total 
emission coefficients
\begin{gather}
S(\omega,T)=\mu^{-1}_0\opn{Re} \int\limits_{\mathcal{A}}
\opn{d}\mathbf{ a}(\mathbf{ r})\cdot\big\langle \mathbf{ E}_{}(\mathbf{ r},\omega)\times 
\mathbf{ B}_{}^{*}(\mathbf{ r},\omega)\big\rangle_T \nonumber\\
=\mathbb{S}_{\alpha^{}_+,\alpha_+'}\big\langle\mathcal{A}^{\Pi}_{\alpha^{}_+}
\mathcal{A}^{\Pi *}_{\alpha_+'}\big\rangle_T\;.
\label{SFlux-calculate}
\end{gather}
They in return depend on the emission coefficients of an isolated sphere as in 
Eq.~(\ref{ExB-Spherical-Coord}) through
\begin{gather}
\mathbb{S}_{\alpha^{}_+,\alpha_+'}=\Lambda^{\dagger}_{\beta_-',\alpha_+} 
\tau^{\dagger}_{\beta_+\beta_-'}\mathbb{I}_{\beta_-\tilde{\beta}_+}
\tau^{}_{\tilde{\beta}_+\tilde{\beta}_-'}\Lambda^{}_{\tilde{\beta}_-',\alpha_+'}
\label{Operator-S}
\end{gather}
with the corresponding surface integral
\begin{gather}
\mathbb{I}_{\beta\beta'}=
\mu^{-1}_0\opn{Re}\int\limits_{\partial V_{pl}}\opn{d}\mathbf{ a}(\mathbf{ r})
\cdot{\bs\phi}^{}_{\beta}(\mathbf{ k}_0,\mathbf{ r})\times\big[\nabla\times
{\bs\phi}^{*}_{\beta'}(\mathbf{ k}_0,\mathbf{ r})\big].
\end{gather}
This integral can be solved analytically and only the diagonal terms with
$\beta=\beta'$ do not vanish. In the nonlocal case, this simplifies the 
computation significantly as we can use the results from Sec.~\ref{sec:Ex&Huy}. 

The heat transfer in the nonequilibrium case where the half-space is at 
temperature $T_{pl}$ and the sphere at $T_{sp}$ is obtained by applying the 
reciprocity argument \cite{Polder1971}.  The net heat transfer can then be 
written as
\begin{gather}
Q(T_{pl},T_{sp})=Q(T_{pl},0)+Q(0,T_{sp}) \nonumber\\
=\int\frac{\opn{d}\omega}{2\pi}\;\big[S(\omega,T_{pl})-S(\omega,T_{sp})\big] ,
\nonumber\\
=\int\frac{\opn{d}\omega}{2\pi}\;
\big[\Theta(\omega,T_{pl})-\Theta(\omega,T_{sp})\big]\mathcal{T}(\omega).
\end{gather}
where the last equation has been cast into the Landauer-like form 
\cite{Biehs2010,Biehs2016} by introducing the transmissivity  
$\mathcal{T}(\omega)$ which will be of use in the next section.

\subsection{Numerical Results}
\label{subsect: numerics}

We now apply our results to numerically compute the heat transfer rate between 
a ZnSe sphere and half-space using the dielectric function for ZnSe given by 
Eq.~(\ref{Epsilon_ZnSe}). The temperature of the sphere is 
$k_BT_{sp}=\hbar\omega_T$ and that of the plate is taken to be 5$\%$ smaller.

In order to compute $\mathcal{T}(\omega)$ and $S(T)$ we have to solve 
Eq.~(\ref{Conditinal equation total Amplitude}) for the spherical scattering 
amplitudes $\mathcal{A}^{\Pi}_{\alpha_+'}$. To this end, we introduce a cutoff 
at $n=n_{max}$ which we chose such that $n_{max}=8+2.5k_0R+R/d$. 
Finally, we use these result in Eq.~(\ref{SFlux-calculate}) to 
obtain the heat flux spectral density $S(\omega, T)$, from which we obtain $S(T)$ 
by integrating over the frequency with an adaptive Gauss quadrature method.  
For more details on the numerical convergence and scaling properties of this method,
we refere the reader to Ref.~\cite{Otey2011}.

\begin{figure}[ht]
\centering
\includegraphics[width=12.5cm]{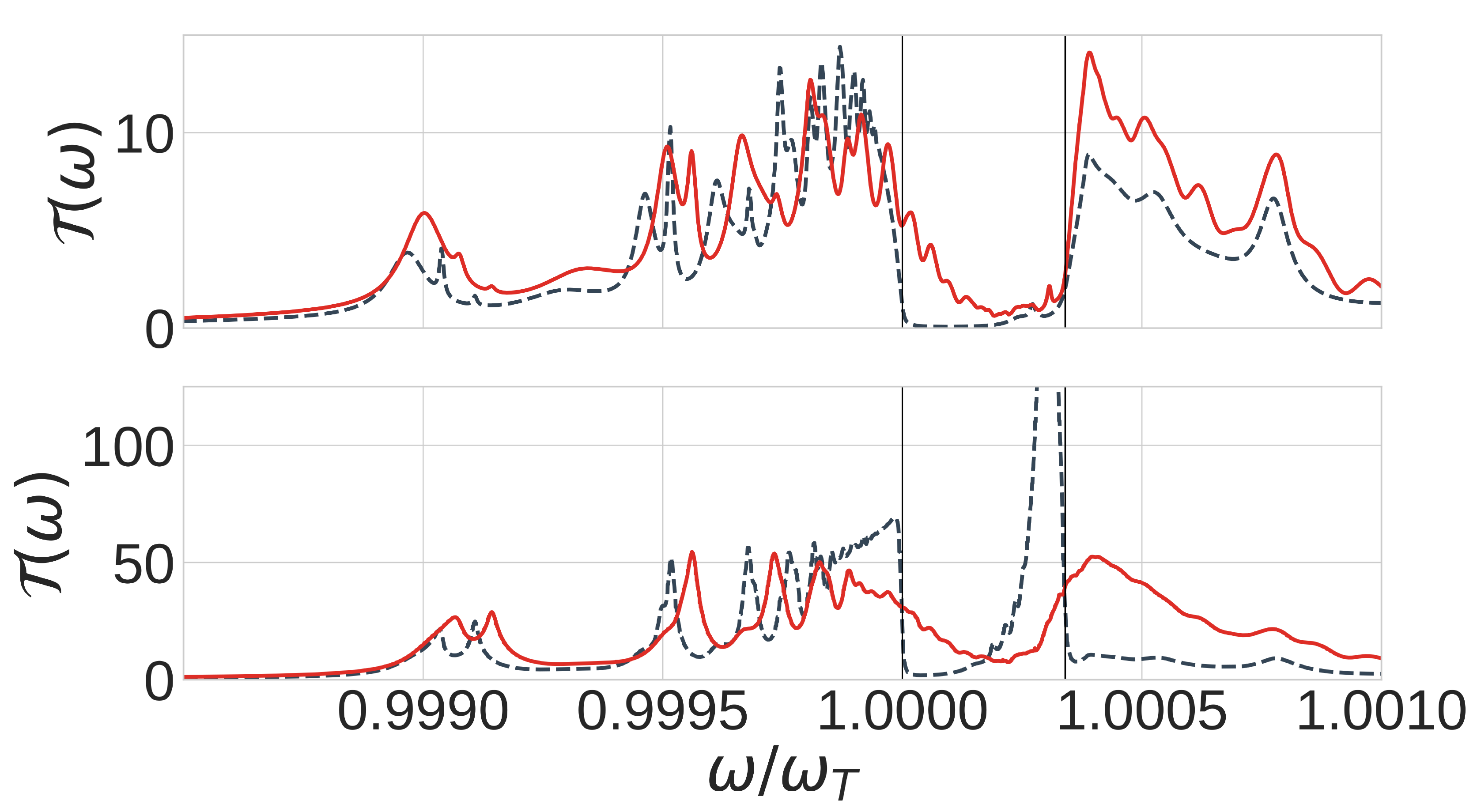}
\caption{\label{fig:transPlSp} Spectral transmissivity for a ZnSe 
sphere with radius $R=220$nm and gap distance to a ZnSe plate of $d=3600$nm 
(top) and $d=1$nm (bottom) with the colour coding: local (dashed, grey) and HuyEx (solid, red).}
\end{figure}

In Fig.~\ref{fig:transPlSp} we show the transmissivity 
$\mathcal{T}(\omega)$ for a sphere of Radius $R=220$ nm as function of 
frequency. In the upper panel, the far-field spectrum is shown for $d=3600nm$, 
roughly eight times the transition wavelength. In the lower panel, the gap 
distance is only $d=1$nm. 

In the far-field spectrum, one observes the expected suppressed transmissivity 
within the stopband $\omega_T<\omega<\omega_L$. In the nonlocal case, this 
effect is enhanced. The previously observed frequency shifts in the emissivity 
are retained. It is apparent that the overall transmissivity in the nonlocal 
case is enhanced. From this, it follows that heat transfer rate is increased as 
well.
In the lower panel of Fig.~\ref{fig:transPlSp} the gap distance is much 
smaller than the transition wavelength $d\ll\lambda_T$. Hence, evanescent modes 
govern the heat transfer rate. Below the transverse resonance frequency 
$\omega_T$, one observes the influence of the whispering gallery modes, and 
within the stop band, the enhancement of the surface guided modes dominate the 
spectrum. In the local case and for $n_{max}=60 $ these modes peak at 
$\mathcal{T}\approx 400$, beyond the shown region. 

When comparing the far-field spectrum with the near-field spectrum, one observes
the enhancement in the transmissivity due to surface guided modes.  We expect 
nonlocal contributions to dominate the resonances whenever the gap distance 
is in the order of $d\sim\sqrt{D 4\pi^2/\omega^2_T}\approx 1.2$nm. This 
is indeed evident from Fig.~\ref{fig:transPlSp}. The surface resonance 
is suppressed, broadened and shifted towards higher frequencies.

\begin{figure}[ht]
\centering
 \includegraphics[width=12.5cm]{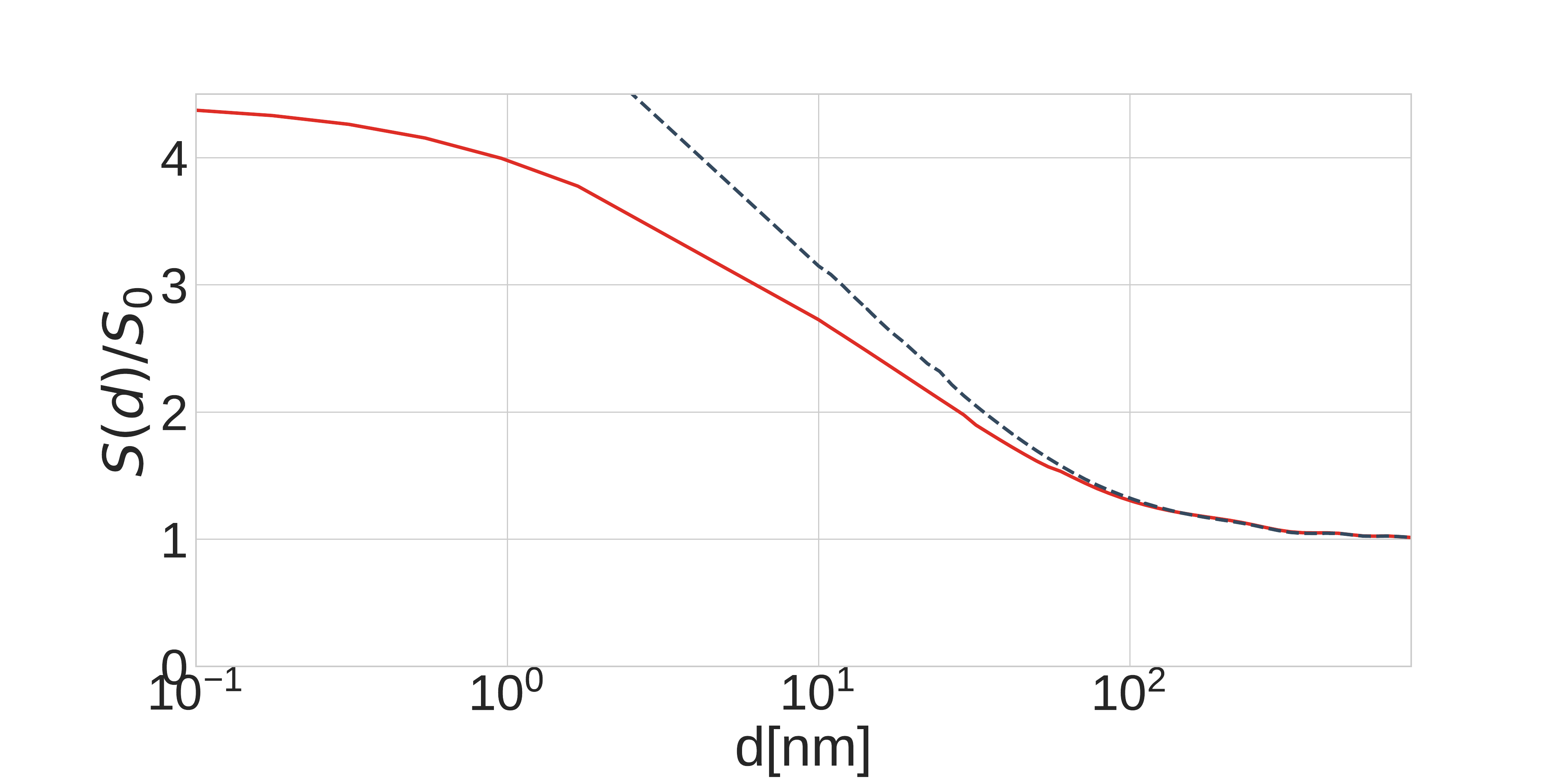}
\caption{\label{fig:sphere-plate-Spect} Heat transfer rate as a 
function of distance. The nonlocal HuyEx (solid, red) results level off at short 
distances, while the local (dashed, grey) case shows the typical spurious divergence 
at short distances.}
\end{figure}

In Fig.~\ref{fig:sphere-plate-Spect} the heat transfer rate $S(d)/S_0$ is shown
as a function of distance, where the normalisation $S_0$ is simply the value of 
$S(d)$ at $d=3600$nm (far-field).
It is worth noting that $S_0$ in the nonlocal case is $36.5\%$ larger than its
local analogue. This is evident from the top panel in Fig.~\ref{fig:transPlSp} 
as the nonlocal spectrum, beyond the transition frequency $\omega_T$, exceeds 
the local spectrum. At short distances $d\lesssim 100$nm, both curves diverge 
from one another, with the result for the spatially dispersive material 
levelling out at a constant value. Hence, the inclusion of spatial 
dispersion removes the divergent behaviour due to the damping of the surface 
guided modes (Fig.~\ref{fig:transPlSp}). The removal of the spurious divergence 
that occurs in a local theory is consistent with recent studies of heat 
transfer between planar boundaries \cite{Churchill2016, Chapuis2008}. However, 
the absolute length scales at which the heat transfer rate tails off is already 
so small that any macroscopic approach could be questionable, and other effects 
might become important.

\section{Summary}\label{sec:Summary} 

We have utilised Huygens' principle and the extinction theorem (HuyEx) to 
derive the reflection coefficients for a spatially dispersive half-space.  We 
found the reflection coefficients to be dependent on surface impedances, which 
can be evaluated for arbitrary homogeneous and isotropic spatially dispersive 
dielectric functions. Numerical results for ZnSe material have shown a 
remarkable resemblance to results obtained with the Fuchs and Kliever ABC's
for most of the frequency and angular spectrum, apart from the longitudinal and 
transverse splitting which, in the vicinity of the longitudinal resonance 
frequency $\omega_L$, is more pronounced for Maxwell boundary conditions.

Based on the fluctuation dissipation theorem, we derived the emissivity of an 
isolated spatially dispersive sphere. We compared the HuyEx and Pekar ABC's to 
the local results and found significant differences for a ZnSe sphere of 
radius $R=220$ nm. The overall emissivity is enhanced in both nonlocal cases.  
The ABC's spectrum shows an increased number of resonances, and in contrast to 
the HuyEx and local case, the maximal emissivity lies within the stop band.  
The HuyEx emissivity is much more similar to the local result. The peaks are 
shifted towards higher frequencies, and additional resonances appear above 
$\omega_T$. 

Finally, we used an exact mode matching method to investigate the influence of 
spatial dispersion on the near-field heat transfer. We found a significant 
impact of spatial dispersion on the spectral heat transfer rate. In particular, 
the surface guided modes are suppressed, blue shifted and broadened compared to 
the local analogue. As a consequence and in contrast to the local case, the heat 
transfer rate levels off at small distances, thereby removing the spurious 
divergences that plague local theories.

\section*{Acknowledgments}
This work was supported by the Deutsche Forschungsgemeinschaft (Collaborative 
Research Center SFB 652/3). 

\appendix
\section{Basis vectors and orthogonality relations}
\label{Appendix:BasisVectors}

\subsection{Vector cylindrical harmonics}
\label{Appendix:VectorCylindricalHarmonics}

The vector wave function in cylindrical coordinates used in this article are 
defined as
\begin{gather}
\mathbf{M}_n(\mathbf{k},\mathbf{r})=
\frac{\mathcal{N}_n}{q}\nabla\times\mathbf{ e}_zJ_n(q\rho)e^{i\beta z+in\phi},
\nonumber\\
\mathbf{N}_n(\mathbf{k},\mathbf{r})=
\frac{\mathcal{N}_n}{q k}\nabla\times\nabla\times
\mathbf{e}_zJ_n(q\rho)e^{i\beta z+in\phi}, \nonumber \\
\mathbf{L}_n(\mathbf{k},\mathbf{r})=\frac{1}{k}\mathcal{N}_n\nabla 
J_n(q\rho)e^{i\beta z +i n \phi}
\end{gather}\label{VectorCyl}
with the normalisation constant $\mathcal{N}_n=i^n/2\pi$. Note that 
$(-1)^n\mathcal{N}_{-n}=\mathcal{N}_n$. The normalisation constants are chosen 
such that the vector cylindrical harmonics are orthonormal, i.e.
\begin{gather}
\int\opn{d}^3r\mathbf{M}_n(\mathbf{k},\mathbf{r})\cdot 
\mathbf{M}_{-n'}(-\mathbf{ k}',\mathbf{r})
=\delta_{n,n'}\delta(\beta-\beta')\frac{\delta(\mathbf{q}-\mathbf{ q}')}{q},
\nonumber\\
\int\opn{d}^3r\mathbf{N}_n(\mathbf{k},\mathbf{r})\cdot 
\mathbf{N}_{-n'}(-\mathbf{ k}',\mathbf{r})
=\delta_{n,n'}\delta(\beta-\beta')\frac{\delta(\mathbf{q}-\mathbf{ q}')}{q},
\nonumber\\
\int\opn{d}^3r\mathbf{L}_n(\mathbf{k},\mathbf{r})\cdot 
\mathbf{L}_{-n'}(-\mathbf{ k}',\mathbf{r})
=\delta_{n,n'}\delta(\beta-\beta')\frac{\delta(\mathbf{q}-\mathbf{ q}')}{q}.
\end{gather}

\subsection{Field expansion}\label{Appendix:FieldExpension}

We require an expansion of the field into components orthogonal to the 
surface and parallel to it. For this purpose we introduce a new orthonormal 
basis
\begin{gather}
\mathbf{ X}_n(\mathbf{q},\phi,\rho)
=\tilde{\mathcal{N}}_nJ_n(q\rho)e^{in\phi}\mathbf{e}_q, \nonumber\\
\mathbf{e}_z\times\mathbf{X}_n(\mathbf{ q},\phi,\rho)
=-\tilde{\mathcal{N}}_nJ_n(q\rho)e^{in\phi}\mathbf{e}_q\times\mathbf{ e}_z, 
\nonumber\\
{\bs \chi}_n(\mathbf{ q},\phi,\rho)
=\tilde{\mathcal{N}}_nJ_n(q\rho)e^{in\phi}\mathbf{e}_z,
\end{gather}\label{angularBasis}
with $\tilde{\mathcal{N}}_n=\frac{i^n}{\sqrt{2\pi}}$. They obey the 
orthogonality relations
\begin{gather}
\int\opn{d}\phi\int\opn{d}\rho\rho \mathbf{X}_{n}(\mathbf{q},\phi,\rho)\cdot 
\mathbf{X}_{-n'}(-\mathbf{q}',\phi,\rho) 
=-\frac{\delta(q-q')}{q}\delta_{nn'} ,
\nonumber\\
\int\opn{d}\phi\int\opn{d}\rho\rho 
\big[\mathbf{e}_z\times\mathbf{X}_{n}(\mathbf{ q},\phi,\rho)\big]\cdot 
\big[\mathbf{e}_z\times\mathbf{X}_{-n'}(-\mathbf{ q}',\phi,\rho)\big]
=-\frac{\delta(q-q')}{q}\delta_{nn'} ,\nonumber\\
\int\opn{d}\phi\int\opn{d}\rho\rho{\bs \chi}_{n}(\mathbf{q},\phi,\rho)\cdot 
{\bs \chi}_{-n'}(-\mathbf{q}',\phi,\rho)=\frac{\delta(q-q')}{q}\delta_{nn'}.
\end{gather}
They are connected to the vector cylindrical harmonics as
\begin{gather}
\mathbf{M}_n(\mathbf{k},\mathbf{r})
=-i\frac{1}{\sqrt{2\pi}}e^{i\beta z}\mathbf{ e}_z\times
\mathbf{X}_{n}(\mathbf{q},\phi,\rho) ,\nonumber\\  
\mathbf{N}_n(\mathbf{k},\mathbf{r})
=\frac{1}{\sqrt{2\pi}}\frac{1}{k}e^{i\beta z}
\big[q{\bs \chi}_n(\mathbf{q},\phi,\rho)-\beta 
\mathbf{X}_{n}(\mathbf{ q},\phi,\rho)\big],\nonumber\\
\mathbf{L}_n(\mathbf{k},\mathbf{r})
=\frac{1}{\sqrt{2\pi}}\frac{i}{k}e^{i\beta z}
\big[\beta{\bs \chi}_n(\mathbf{q},\phi,\rho)+q 
\mathbf{X}_{n}(\mathbf{ q},\phi,\rho)\big].
\end{gather}

\subsection{Important relations and basis transformations}
\label{Appendix:Basetransformations}
In equation \ref{reflxcoef} we derived the reflection coefficients $R_s$ and $R_q$ in the basis of Eq.~(\ref{angularBasis}).
we are interested in the relation between these reflection coefficients and the corresponding reflection coefficients in the basis of vector cylindrical harmonics $r_s$, $r_p$.
Let the field $\mathbf{ F}(\mathbf{r})$ be in the basis of 
Eq.~(\ref{fieldExpansionPlate}). Then,
\begin{gather}
\alpha_n(\mathbf{q},\beta)
=\int\opn{d}^2r \mathbf{ F}(\mathbf{ r})\cdot
\mathbf{M}_{-n}(-\mathbf{q},-\beta,\mathbf{r}) \nonumber\\
=\sum_{n'}\int\opn{d}z \int\opn{d}q\;q A_{n'}(\mathbf{q},z) 
i\frac{e^{-i\beta z}}{\sqrt{2\pi}}\frac{\delta(q-q')}{q}\delta_{nn'}.
\end{gather}
For the incoming field we may choose
\begin{gather}
\alpha^{inc}_n(\mathbf{q},\beta) 
=\alpha^{inc}_n(\mathbf{ q}_0,\beta_0)\frac{\delta(q-q_0)}{q}\delta(\beta+\beta_0) 
\nonumber \\
=\int\opn{d}z A^{inc}_{n}(\mathbf{q},z)i
\frac{e^{-i\beta z}}{\sqrt{2\pi}}.
\end{gather}
The term $\delta(\beta+\beta_0)$ takes the direction of the incoming wave, 
traveling into the $-z$ direction, into account. We can also derive 
$A^{inc}_{n}(\mathbf{q},z)$ in terms of $\alpha^{inc}_n(\mathbf{q},\beta)$ and 
find
\begin{gather}
A^{inc}_{n}(\mathbf{q},z)
=-\int\opn{d}^3r \mathbf{ F}(\mathbf{r})\cdot\mathbf{ e}_z\times
\mathbf{X}_{-n}(-\mathbf{q},-\beta,\mathbf{ \rho}) \nonumber\\
=-\frac{i}{\sqrt{2\pi}}\int\opn{d}\beta e^{i\beta z}\alpha_n(\mathbf{q},\beta) 
,\nonumber\\
A^{inc}_{n}(\mathbf{q},z)
=-\frac{i}{\sqrt{2\pi}}\alpha^{inc}_n(\mathbf{ q}_0,\beta_0)\frac{\delta(q-q_0)}{q}
\frac{e^{-i\beta_0 z}}{\sqrt{2\pi}}.
\end{gather}
In order to expand the reflection amplitudes in terms of reflection 
coefficients,
\begin{gather}
A^{scat}_n(\mathbf{q},z)=A^{inc}_n(\mathbf{q},0)R_s(\mathbf{q},z),\nonumber
\label{Temp2}
\end{gather}
we decompose $\alpha^{scat}_n(\mathbf{q},\beta)$ by 
writing $\alpha^{scat}_n(\mathbf{q},\beta)=r_s(\mathbf{ 
q},\beta)\alpha^{inc}_{n}(\mathbf{q},-\beta)$, taking into account that the 
scattered field travels in the opposite $z$ direction to the incoming field. 
Thus, we find
\begin{gather}
 A^{scat}_{n}(\mathbf{q},z)
 =A^{inc}_n(\mathbf{q},0)R_s(\mathbf{q},z)= 
\frac{i}{\sqrt{2\pi}}\int\opn{d}\beta e^{i\beta z}\alpha_n(\mathbf{q},\beta) 
\nonumber \\
 =\frac{i}{\sqrt{2\pi}}r_s(\mathbf{q},\beta_0)
 \alpha^{inc}_n(\mathbf{ q}_0,\beta_0)\frac{\delta(q-q_0)}{q}
 \frac{e^{i\beta_0 z}}{\sqrt{2\pi}}.
\end{gather}
Using the above and $R_s(\mathbf{q},z)=R_s(\mathbf{q})e^{i\beta z}$ we 
obtain 
\begin{gather}
R_s(\mathbf{q}_0)=r_s(\mathbf{q}_0,\beta_0).
\end{gather}
 
Similarly, we can relate $R_q$ with $r_p$. Therefore we need to take into account 
that the incoming and reflected field are transversal only. Hence $\gamma_n$ in 
Eq.~(\ref{MNL(k)-basis}) vanishes. As a consequence $B^{(0)}_n$ and $C^{(0)}_n$ are linear dependent. 
In analogy to the above procedure, we find
\begin{gather}
r_p(\mathbf{q}_0,\beta_0)=-R_q(\mathbf{q}_0).
\end{gather}

\section{Algebraic transformation of the fluctuation-dissipation theorem}
\label{Appendix:ExB}

In this appendix we derive Eq.~(\ref{Poitning TheoremNLSphere}). The starting 
point is Eq.~(\ref{ExB}) together with the fluctuation-dissipation 
theorem Eq.~(\ref{FDT}),
\begin{gather}
\big\langle \mathbf{E}(\mathbf{r},\omega)\times 
\mathbf{ B}^{*}(\mathbf{ r}',\omega)\big\rangle_T 
=-2\epsilon_0\mu^3_0\omega^3\Theta(\omega,T)\int\limits_
{V_{\opn{sp}}}\opn{d}^3s\int\limits_{V_{\opn{sp}}}\opn{d}^3s'
\opn{O_{c}}\big\lbrace 
\ten{G}(\mathbf{r},\mathbf{ s})\cdot
\opn{Im}\ten{\epsilon}(\mathbf{ s},\mathbf{ s}',\omega)\cdot
\mathbf{\Gamma}^{\dagger}(\mathbf{r}',\mathbf{ s}') 
\big\rbrace.
\label{fdt}
\end{gather}
If we split the imaginary part of the dielectric tensor into 
$\opn{Im}\ten{\epsilon}(\mathbf{ s},\mathbf{ s}')
=(2i)^{-1}\big[\ten{\epsilon}(\mathbf{ s},\mathbf{ s}')
-\ten{\epsilon}^{*}(\mathbf{ s},\mathbf{ s}')\big]$,
and use the Helmholtz equation for the Green tensor, we can eliminate the 
dielectric tensor from Eq.~(\ref{fdt}) and find
\begin{gather}
\big\langle \mathbf{E}(\mathbf{r},\omega)\times 
\mathbf{ B}^{*}(\mathbf{ r}',\omega)\big\rangle_T
=i\mu^2_0\omega\Theta(\omega,T)\opn{O_{c}} \bigg\lbrace \nonumber\\ 
\int\limits_{ V_{\opn{sp}}}\opn{d}^3s'\big[
\ten{G}(\mathbf{r},\mathbf{ s}')\times\overleftarrow{\nabla}_{s'}
\times\overleftarrow{\nabla}_{s'}-\ten{I} 
\delta(\mathbf{r}-\mathbf{ s}')\big]\cdot
\mathbf{ \Gamma}^{\dagger}(\mathbf{r}',\mathbf{ s}') \nonumber\\
-\int\limits_{V_{\opn{sp}}}\opn{d}^3s\,\ten{G}(\mathbf{r},\mathbf{ s})\cdot\big[
\nabla_s\times\nabla_s\times
\mathbf{ \Gamma}^{\dagger}(\mathbf{r}',\mathbf{ s}) 
-\frac{\ten{I} \delta(\mathbf{ s}-\mathbf{ r}')}{i\mu_0\omega}\times
\overleftarrow{\nabla}_{r'}\big]\bigg\rbrace.
\end{gather}
Note that both integrals over $s$ and $s'$ are finite volume integrals and the 
points $\mathbf{r}$ and $\mathbf{r}'$ are located outside this volume. Thus, 
the delta function terms do not contribute to the integrals. Hence, we can write
\begin{gather}
\big\langle \mathbf{E}(\mathbf{r},\omega)\times 
\mathbf{ B}^{*}(\mathbf{ r}',\omega)\big\rangle_T
=i\mu^2_0\omega\Theta(\omega,T) 
\opn{O_{c}} \bigg\lbrace\int\limits_{V_{\opn{sp}}}  \opn{d}^3s\bigg[ 
\ten{G}(\mathbf{r},\mathbf{ s})\times\overleftarrow{\nabla}_{s}\times
\overleftarrow{\nabla}_{s}\cdot\mathbf{ \Gamma}^{\dagger}(\mathbf{ r'},\mathbf{ s})
\nonumber\\
-\ten{G}(\mathbf{r},\mathbf{ s})\cdot\big[\nabla_s\times\nabla_s\times
\mathbf{ \Gamma}^{\dagger}(\mathbf{r}',\mathbf{ s})\big]\bigg]\bigg\rbrace\;.
\end{gather}

It is convenient to eliminate the outer curl and transform the volume
integral into a surface integral by applying the vector Green theorem. For this 
reason we first transform the outer vector product into an outer product
\begin{gather}
\opn{O_{c}}\big\lbrace \big(\big(\ten{G}(\mathbf{r},\mathbf{ s})
\times\overleftarrow{\nabla}_{s}\big)\times\overleftarrow{\nabla}_{s}
\big)\cdot\mathbf{ \Gamma}^{\dagger}(\mathbf{ r'},\mathbf{ s})\big\rbrace
=\opn{O_{p}}\big\lbrace\nabla_s\times\nabla_s\times
\ten{G}^{T}(\mathbf{ r},\mathbf{ s})\times\mathbf{ \Gamma}^{*}(\mathbf{r}',\mathbf{ s}) 
\big\rbrace, \nonumber\\
\opn{O_{c}}\big\lbrace \ten{G}(\mathbf{r},\mathbf{ s})\cdot\big( 
\nabla_{s}\times\big(\nabla_{s}\times
\mathbf{ \Gamma}^{\dagger}(\mathbf{r}',\mathbf{ s})\big)\big)\big\rbrace 
=\opn{O_{p}}\big\lbrace\ten{G}^{T}(\mathbf{r},\mathbf{ s})\times\big
(\mathbf{ \Gamma}^{*}(\mathbf{r}',\mathbf{ s})\times
\overleftarrow{\nabla}_s\big)\times\overleftarrow{\nabla}_s\big\rbrace\;.
\end{gather}
This can be applied to
\begin{gather}
\nabla_s\cdot\opn{O_{c}}\big\lbrace 
\big[\nabla_s\times\ten{G}^{T}(\mathbf{r},\mathbf{ s})\big]\times
\mathbf{ \Gamma}^{*}(\mathbf{r}',\mathbf{ s})\big\rbrace 
= \opn{O_{p}}\big\lbrace\nabla_s\times\big(\nabla_s\times
\ten{G}^{T}(\mathbf{r},{ \bf s})\big)\times
\mathbf{ \Gamma}^{*}(\mathbf{r}',\mathbf{ s})\big\rbrace \nonumber\\
+\opn{O_{p}}\big\lbrace\nabla_s\times
\ten{G}^T(\mathbf{r},\mathbf{ s})\times\big(
\mathbf{ \Gamma}^{*}(\mathbf{r}',\mathbf{ s})\times
\overleftarrow{\nabla}_s\big)\big\rbrace,\nonumber\\
\nabla_s\cdot\opn{O_{c}}\big\lbrace 
\ten{G}^{T}(\mathbf{r},\mathbf{ s})\times\big[
\mathbf{ \Gamma}^{*}(\mathbf{r}',\mathbf{ s})\times
\overleftarrow{\nabla}_s\big]\big\rbrace
=\opn{O_{p}}\big\lbrace\nabla_s\times\ten{G}^{T}(\mathbf{r},\mathbf{ s})\times
\mathbf{ \Gamma}^{*}(\mathbf{r}',\mathbf{ s})\times\overleftarrow{\nabla}_s\big\rbrace
\nonumber\\
+\opn{O_{p}}\big\lbrace\ten{G}^{T}(\mathbf{r},\mathbf{ s})\times\big(
\mathbf{ \Gamma}^{*}(\mathbf{r}',\mathbf{ s})\times\overleftarrow{\nabla}_s\big)
\times\overleftarrow{\nabla}_s\big\rbrace\; .
\end{gather}
Hence we can rewrite
\begin{gather}
\opn{O_{c}}\big\lbrace \big(\big(\ten{G}(\mathbf{r},\mathbf{ s})
\times\overleftarrow{\nabla}_{s}\big)\times\overleftarrow{\nabla}_{s}
\big)\cdot\mathbf{ \Gamma}^{\dagger}(\mathbf{ r'},\mathbf{ s})\big\rbrace 
-\opn{O_{c}}\big\lbrace \ten{G}(\mathbf{r},\mathbf{ s})\cdot\big( 
\nabla_{s}\times\big(\nabla_{s}\times\mathbf{ \Gamma}^{\dagger}(\mathbf{r}',\mathbf{ s})
\big)\big)\big\rbrace \nonumber\\
=\opn{O_{p}}\big\lbrace\nabla_s\times\nabla_s\times
\ten{G}^{T}(\mathbf{r},\mathbf{ s})
\times\mathbf{ \Gamma}^{*}(\mathbf{r}',\mathbf{ s})\big\rbrace 
-\opn{O_{p}}\big\lbrace\ten{G}^{T}(\mathbf{r},\mathbf{ s})\times\big(
\mathbf{ \Gamma}^{*}(\mathbf{r}',\mathbf{ s})\times\overleftarrow{\nabla}_s\big)
\times\overleftarrow{\nabla}_s\big\rbrace \nonumber\\
=\nabla_s\cdot\opn{O_{c}}\big\lbrace 
\big[{\nabla}_s\times\ten{G}^{T}(\mathbf{r},\mathbf{ s})\big]\times
\mathbf{ \Gamma}^{*}(\mathbf{r}',\mathbf{ s})\big\rbrace 
-\nabla_s\cdot\opn{O_{c}}\big\lbrace 
\ten{G}^{T}(\mathbf{ s},\mathbf{r})\times\big[
\mathbf{ \Gamma}^{*}(\mathbf{r}',\mathbf{ s})\times\overleftarrow{\nabla}_s
\big]\big\rbrace.
\end{gather}
Thus we find
\begin{gather}
\big\langle \mathbf{E}(\mathbf{r},\omega)\times 
\mathbf{ B}^{*}(\mathbf{r}',\omega)\big\rangle_T 
=i\mu^2_0\omega\Theta(\omega,T) 
\int\limits_{\partial V}\opn{d}^2 
s\;\opn{O_{p}}\Big\lbrace\big[\mathbf{ n}(\mathbf{ s})\times
\ten{G}^{T}(\mathbf{r},\mathbf{ s})\big]\times\big[
\mathbf{ \Gamma}^{*}(\mathbf{r}',\mathbf{ s})\times\overleftarrow{\nabla}_s\big] 
\nonumber\\
-\mathbf{ n}(\mathbf{ s})\times\big[\nabla_s\times\ten{G}^{T}(\mathbf{r},\mathbf{ s})
\big]\times\mathbf{ \Gamma}^{*}(\mathbf{r}',\mathbf{ s})\Big\rbrace
\end{gather}
from which Eq.~(\ref{Poitning TheoremNLSphere}) follows.\\


%
\end{document}